\documentclass[acmsmall,screen,nonacm,authoryear]{acmart} 
\AtBeginDocument{%
  }

\usepackage{caption}
\usepackage{xcolor}
\usepackage{array} 
\usepackage{pifont}
\usepackage{listings}
\usepackage{amsmath}
\usepackage{mathpartir,amsmath}
\usepackage{graphicx}
\usepackage{subcaption}

\usepackage{tikz}
\usetikzlibrary{arrows.meta,positioning,calc,matrix}
\setcopyright{acmlicensed}
\copyrightyear{2018}
\acmYear{2018}
\acmDOI{XXXXXXX.XXXXXXX}

\acmJournal{JACM}
\acmVolume{37}
\acmNumber{4}
\acmArticle{111}
\acmMonth{8}

\newcommand{\bnfdef}{\mathrel{::=}}
\newcommand{\bnfalt}{\mathrel{\mid}}

\usepackage{xcolor}
\definecolor{Ared}{RGB}{220,40,40}
\definecolor{Bblue}{RGB}{40,80,200}
\definecolor{Cgreen}{RGB}{20,150,70}
\newcommand{\Aio}[1]{\textcolor{Ared}{#1}}
\newcommand{\Bio}[1]{\textcolor{Bblue}{#1}}
\newcommand{\Cio}[1]{\textcolor{Cgreen}{#1}}

\newcommand{\firstblock}{%
\scriptsize
\textit{Cell $(0,0)$.}\\[2pt]\noindent
\begin{flushleft}
$\displaystyle
\begin{array}{@{}l@{}}
\Aio{[i,k]_W \leftarrow A_{ik}} \\[3pt]
\text{for } j \\[3pt]
\; \Bio{[i,k]_B \leftarrow B_{kj}} \\
\;[i,k]_C \leftarrow 0 + [i,k]_W \cdot [i,k]_B \\[3pt]
\;[i,k]_B \rightarrow [i+1,k]_B\\
\;[i,k]_C \rightarrow [i,k+1]_C\\
\end{array}
$
\end{flushleft}
}
\newcommand{\firstcell}{%
  \fbox{%
    \begin{minipage}[t]{0.31\linewidth}\raggedright
      \firstblock
    \end{minipage}%
  }%
}

\newcommand{\secondblock}{%
\scriptsize
\textit{Cell $(0,k)$.}\\[2pt]\noindent
\begin{flushleft}
$\displaystyle
\begin{array}{@{}l@{}}
\Aio{[i,k]_W \leftarrow A_{ik}} \\[3pt]
\text{for } j \\[3pt]
\;\Bio{[i,k]_B \leftarrow B_{kj}} \\
\;[i,k]_C \leftarrow [i,k\!-\!1]_C + [i,k]_W \cdot [i,k]_B \\[3pt]
\;[i,k]_B \rightarrow [i+1,k]_B\\
\;[i,k]_C \rightarrow [i,k+1]_C\\
\end{array}
$
\end{flushleft}
}
\newcommand{\secondcell}{%
  \fbox{%
    \begin{minipage}[t]{0.31\linewidth}\raggedright
      \secondblock
    \end{minipage}%
  }%
}

\newcommand{\thirdblock}{%
\scriptsize
\textit{Cell $(0,k_{end})$.}\\[2pt]
\noindent
\begin{flushleft}
$\displaystyle
\begin{array}{@{}l@{}}
\Aio{[i,k]_W \leftarrow A_{ik}} \\[3pt]
\text{for } j \\[3pt]
\;\Bio{[i,k]_B \leftarrow B_{kj}} \\
\;[i,k]_C \leftarrow [i,k\!-\!1]_C + [i,k]_W \cdot [i,k]_B \\[3pt]
\;[i,k]_B \rightarrow [i+1,k]_B\\
\;\Cio{C_{ij} \leftarrow [i,k]_C}\\
\end{array}
$
\end{flushleft}
}
\newcommand{\thirdcell}{%
  \fbox{%
    \begin{minipage}[t]{0.31\linewidth}\raggedright
      \thirdblock
    \end{minipage}%
  }%
}

\newcommand{\fourthblock}{%
\scriptsize
\textit{Cell $(i,0)$.}\\[2pt]
\noindent
\begin{flushleft}
$\displaystyle
\begin{array}{@{}l@{}}
\Aio{[i,k]_W \leftarrow A_{ik}} \\[3pt]
\text{for } j \\[3pt]
\;[i,k]_B \leftarrow [i\!-\!1,k]_B \\
\;[i,k]_C \leftarrow 0 + [i,k]_W \cdot [i,k]_B \\[3pt]
\;[i,k]_B \rightarrow [i+1,k]_B\\
\;C_{ij} \leftarrow [i,k]_C\\
\end{array}
$
\end{flushleft}
}
\newcommand{\fourthcell}{%
  \fbox{%
    \begin{minipage}[t]{0.31\linewidth}\vspace{0pt}%
      \fourthblock
    \end{minipage}%
  }%
}

\newcommand{\fifthblock}{%
\scriptsize
\textit{Cell $(i,k)$.}\\[2pt]
\noindent
\begin{flushleft}
$\displaystyle
\begin{array}{@{}l@{}}
\Aio{[i,k]_W \leftarrow A_{ik}} \\[3pt]
\text{for } j \\[3pt]
\;[i,k]_B \leftarrow [i\!-\!1,k]_B \\
\;[i,k]_C \leftarrow [i,k\!-\!1]_C + [i,k]_W \cdot [i,k]_B \\[3pt]
\;[i,k]_B \rightarrow [i+1,k]_B\\
\;[i,k]_C \rightarrow [i,k+1]_C\\
\end{array}
$
\end{flushleft}
}
\newcommand{\fifthcell}{%
  \fbox{%
    \begin{minipage}[t]{0.31\linewidth}\vspace{0pt}%
      \fifthblock
    \end{minipage}%
  }%
}

\newcommand{\sixthblock}{%
\scriptsize
\textit{Cell $(i,k_{end})$.}\\[2pt]
\noindent
\begin{flushleft}
$\displaystyle
\begin{array}{@{}l@{}}
\Aio{[i,k]_W \leftarrow A_{ik}} \\[3pt]
\text{for } j \\[3pt]
\;[i,k]_B \leftarrow [i\!-\!1,k]_B \\
\;[i,k]_C \leftarrow [i,k\!-\!1]_C + [i,k]_W \cdot [i,k]_B \\[3pt]
\;[i,k]_B \rightarrow [i+1,k]_B\\
\;\Cio{[C_{ij} \leftarrow [i,k]_C}\\
\end{array}
$
\end{flushleft}
}
\newcommand{\sixthcell}{%
  \fbox{%
    \begin{minipage}[t]{0.31\linewidth}\vspace{0pt}%
      \sixthblock
    \end{minipage}%
  }%
}

\newcommand{\seventhblock}{%
\scriptsize
\textit{Cell $(i_{end},0)$.}\\[2pt]
\noindent
$$\begin{array}{@{}l@{}}
\Aio{[i,k]_W \leftarrow A_{ik}} \\[3pt]
\text{for } j \\[3pt]
\;[i,k]_B \leftarrow [i\!-\!1,k]_B \\
\;[i,k]_C \leftarrow 0 + [i,k]_W \cdot [i,k]_B \\[3pt]
\\
\;[i,k]_C \rightarrow [i,k+1]_C\\
\end{array}$$
}
\newcommand{\seventhcell}{%
  \fbox{%
    \begin{minipage}[t]{0.31\linewidth}\vspace{0pt}%
      \seventhblock
    \end{minipage}%
  }%
}

\newcommand{\eigthblock}{%
\scriptsize
\textit{Cell $(i_{end},k)$.}\\[2pt]
\noindent
$$\begin{array}{@{}l@{}}
\Aio{[i,k]_W \leftarrow A_{ik}} \\[3pt]
\text{for } j \\[3pt]
\;[i,k]_B \leftarrow [i\!-\!1,k]_B \\
\;[i,k]_C \leftarrow [i,k\!-\!1]_C + [i,k]_W \cdot [i,k]_B \\[3pt]
\\
\;[i,k]_C \rightarrow [i,k+1]_C\\
\end{array}$$
}
\newcommand{\eigthcell}{%
  \fbox{%
    \begin{minipage}[t]{0.31\linewidth}\vspace{0pt}%
      \eigthblock
    \end{minipage}%
  }%
}

\newcommand{\ninthblock}{%
\scriptsize
\textit{Cell $(i_{end},k_{end})$.}\\[2pt]
\noindent
$$\begin{array}{@{}l@{}}
\Aio{[i,k]_W \leftarrow A_{ik}} \\[3pt]
\text{for } j \\[3pt]
\;[i,k]_B \leftarrow [i\!-\!1,k]_B \\
\;[i,k]_C \leftarrow [i,k\!-\!1]_C + [i,k]_W \cdot [i,k]_B \\[3pt]
\\
\;\Cio{C_{ij} \leftarrow [i,k]_C}\\
\end{array}$$
}
\newcommand{\ninthcell}{%
  \fbox{%
    \begin{minipage}[t]{0.31\linewidth}\vspace{0pt}%
      \ninthblock
    \end{minipage}%
  }%
}

\begin{document}
\citestyle{acmauthoryear}

\title{Cyclotron: Compilation of Recurrences to Distributed and Systolic Architectures}

\author{Shiv Sundram}
\email{shiv1@stanford.edu}
\affiliation{%
  \institution{Stanford University}
  \city{Stanford}
  \state{California}
  \country{USA}
}

\author{Akhilesh Balasingam}
\email{avb03@stanford.edu}
\affiliation{%
  \institution{Stanford University}
  \city{Stanford}
  \state{California}
  \country{USA}
}

\author{Nathan Zhang}
\email{stanfurd@stanford.edu}
\affiliation{%
  \institution{Stanford University}
  \city{Stanford}
  \state{California}
  \country{USA}
}

\author{Kunle Olukotun}
\email{kunle@stanford.edu}
\affiliation{%
  \institution{Stanford University}
  \city{Stanford}
  \state{California}
  \country{USA}
}

\author{Fredrik Kjolstad}
\email{kjolstad@stanford.edu}
\affiliation{%
  \institution{Stanford University}
  \city{Stanford}
  \state{California}
  \country{USA}
}

\renewcommand{\shortauthors}{Trovato et al.}

\definecolor{commentgreen}{RGB}{2,112,10}
\definecolor{weborange}{RGB}{255,165,0}
\definecolor{frenchplum}{RGB}{129,20,83}

\newcommand*\colourcheck[1]{%
  \expandafter\newcommand\csname #1check\endcsname{\textcolor{#1}{\ding{52}}}%
}

\colourcheck{commentgreen}
\newcommand{\F}{{\commentgreencheck}}
\begin{abstract}
We present Cyclotron, a framework and compiler for using recurrence equations to express streaming dataflow algorithms, which then get portably compiled to distributed topologies of interlinked processors. Our framework provides an input language of recurrences over logical tensors, which then gets lowered into an intermediate language of recurrences over logical iteration spaces, and finally into programs of send, receive, and computation operations specific to each individual processor. In Cyclotron’s IR, programs are optimized such that external memory interactions are confined to the boundaries of the iteration space. Within inner iteration spaces, all data accesses become local: data accesses target values residing in local fast memory or on neighboring processing units, avoiding costly memory movement. We provide a scheduling language allowing users to define how data gets streamed and broadcasted between processors, enabling pipelined execution of computation kernels over distributed topologies of processing elements. We demonstrate the portability of our approach by compiling our IR to a reconfigurable simulator of systolic arrays and chiplet-style distributed hardware, as well as to distributed-memory CPU clusters. In the simulated reconfigurable setting, we use our compiler for hardware design space exploration in which link costs and latencies can be specified. In the distributed CPU setting, we show how to use recurrences and our scheduling language to express various matrix multiplication routines (Cannon, SUMMA, PUMMA, weight stationary) and solvers (Triangular solve and Cholesky). For matrix multiplication and the triangular solve, we generate distributed implementations competitive with ScaLAPACK. 
\end{abstract}



\maketitle

\section{Introduction}
Generations of architectural improvements have augmented the performance gap between data movement and computation. Performance optimization is now increasingly concerned with optimizing memory movement and network traffic, and less on optimizing and reducing the number of floating operations performed. Modern architectures, whether it be a single GPU/accelerator, or a cluster of CPUs, share a common motif of being an interconnected grid of individual processing elements (PEs), arranged either in a lattice or a tree of wires.

These architectures emphasize the need to keep data close to processing elements, and to move data exclusively amongst neighbors, thus avoiding expensive accesses to offchip memory. Therefore, optimized kernels, like those for matrix multiply, are now designed as programs where each timestep consists of three phases. A \emph{receive} phase to import data from nearest neighbors. A \emph{compute} phase to operate on this data. And a final \emph{send} phase to send data to neighboring PEs for the subsequent timestep. Such a design allows all data movement to take the form of less expensive nearest neighbor exchanges (which can often be overlapped with the computation), ameliorating the need for deeper memory movement across machine hierarchies.  

The majority of work in this space has focused on optimizing matrix multiply. In a distributed memory setting, communication-avoiding matrix multiply algorithms like Cannon’s, SUMMA, and PUMMA all fit within this three-phase motif. On a single chip setting, this has manifested in the development of systolic architectures, like the Tensor Processing Unit, in which a grid of multiply accumulator units collectively compute matrix products, with a similar dataflow to Cannon’s algorithm, but where all the PEs are on a single chip, as opposed to being different CPUs entirely. 

However, we lack a general framework for co-designing both a portable programming language and an architectural design space framework to encapsulate the full space of these algorithms, which reaches far beyond matrix multiply. Table ~\ref{tab:features} illustrates the state of current frameworks for such algorithms. Systems like DISTAL \citep{distal} and architectural design space frameworks like GEMMINI \citep{genc2021gemmini}, for example, focus on distributed matrix multiply operations. Furthermore, routines like triangular solve, the Cholesky Decomposition, and  Flash Attention, all exhibit similar dataflow behavior to matrix multiply, and all fit in a general language of recurrences. They differ in terms of what is computed on each PE, which data is sent on the wires, and in which directions. These computations can have different data dependencies, which must be considered when scheduling computation and data movement. Nevertheless, they fall within this three phase pipeline. 

A common framework and programming languages for dataflow computations, would allow us to describe all such computations succinctly in this language, while making it easy to explore the design space of each computation itself.  The affine recurrence model and the Warp processor \citep{lam2004data} demonstrated this idea in the context of hardware-oriented systolic compilation, providing a tool-chain for mapping imperative and affine loop nests onto a specific processor template. However, their compilation model was tightly coupled to a fixed hardware target, making it difficult to generalize beyond that setting or to distributed implementations. Other mathematical frameworks, like the polyhedral model \citep{FEAUTRIER1991}, and the systolic synthesis framework of Rajopadhye and Fujimoto \citep{rajopadhye1990synthesizing}, provide a mathematical foundation for dependency management and software pipelining. However, they are largely theoretical,they do not show how to generate processor-level code. They rely on successive user-specified linear transformations to schedule and optimize computation, an assumption unnecessary for practically compiling and optimizing recurrences onto distributed architectures.

We show how to describe all these dataflow algorithms in a language of recurrence equations, which along with our scheduling language, embed the dependencies, data-reuse, and dataflow of the underlying computations.  We also show how the schedule is needed to add essential information  on how to specify streaming behavior of the computations. These languages are then compilable to both an MPI based distributed memory architecture, where we get competitive performance with handwritten libraries, as well as a single chip dataflow simulator. 

\newcolumntype{P}[1]{>{\raggedright\arraybackslash}p{#1}}

\begin{table}
\scriptsize
\centering
\caption{Features of different programming frameworks for recurrences and matrix multiplies. A \F denotes the framework contains that feature.}
\label{tab:features}
\begin{tabular}{c|c|c|c|c|c|c|c}
Framework & Space-time mappings & Multi-node & Chiplet/simulator& Matmul & Recurrences & Scheduling Language   \\
\hline
GEMMINI & \F &  & \F & \F &  &  \\
DISTAL &  & \F &  & \F &  & \F \\
Rajopadhyaye et al. & \F  &  &  & \F & \F &  \\
Recuma  &  &  &  & \F & \F & \F  \\
AutoSA & \F &  & \F & \F &  &  \\
Cyclotron (this work) & \F & \F & \F & \F & \F  & \F  \\
\end{tabular}
\end{table}

\section{Motivating Example}

To demonstrate how recurrences can elegantly describe a computation, its data reuse, dataflow, and local communication, we use a matrix multiply and triangular solve as our running examples; each illustrates how to handle recurrences without dependencies and recurrences with dependencies, respectively. Figure \ref{fig:bothRecs} contrasts the two recurrences: matrix multiply, whose outputs are independent and can be computed in parallel, and triangular solve, whose outputs depend on previously computed values. In the latter, already calculated entries of output tensor $X$ are reused to compute future ones.


\begin{figure}[t]
\centering
\begin{minipage}[t]{0.45\textwidth}
\centering
\textbf{Matrix Multiplication (Tensor Algebra)}\\[4pt]
\[
C_{ij} = \sum_{k} A_{ik} \, B_{kj}
\]
\[
\text{or equivalently } C_{ij} = A_{ik} B_{kj}
\]
\end{minipage}
\hfill
\begin{minipage}[t]{0.45\textwidth}
\centering
\textbf{Triangular Solve with Multiple Right Hand Sides (Recurrence Form)}\\[4pt]
\[
X_{ri} = \frac{1}{L_{ii}}
\Bigl(B_{ri} - \sum_{k < i} L_{ij} X_{rj}\Bigr)
\]
\[
\text{where } j<i  
\]
\end{minipage}
\caption{Left: Tensor algebra for matrix multiplication, which contains no dependencies between outputs. Right: Recurrence relation for forward triangular solve, which contains output dependencies in which  $X_{ri}$ is dependent on $X_{rj}$ where $j<i$}
\label{fig:bothRecs}
\end{figure}

 Recurrences are equations involving mathematical operations over indexed tensors.  Representative examples of such recurrences are shown in Figure~\ref{fig:recList}, and the grammar used to describe them formally is shown in Figure~\ref{fig:grammar}. They form a superset over the language of Tensor Algebra, which is a language of elementwise operations and summations over indexed tensors. The typical Tensor Algebra form of a matrix multiply is simply $C_ij = \sum_{k} A_{ik}B_{kj}$. Recurrences extend tensor algebra with additional constructs
\begin{enumerate}
    \item Constraints, in which iteration variables (i.e. $i$ and $j$) can be related to each other via inequalities (e.g., $j<i$ in the triangular solve in Figure ~\ref{fig:bothRecs})
    \item Dependencies among outputs, in which the result tensor appears on the left and right side of the equation, meaning output values are needed to calculate future outputs (e.g. $X_{ij}$ in the triangular solve depends on all $X_{rj} $ where $j<i$. 

\end{enumerate}

Both the matrix multiply recurrence and the triangular solve recurrence are declarative equations that show how to mathematically compute the outputs, given the dependencies. Both recurrences, when augmented with a schedule denoting a desired loop ordering (e.g., $ijk$, $jki$ etc. for matrix multiply), can be lowered to imperative programs of loops and compute statements. The RECUMA \citep{recuma} and REPTILE \citep{reptile}
showed how user-provided recurrences, along with user-provided schedules can be lowered to C code that runs on a single processor and achieves performance parity with hand-optimized code. In the generated C code, tensors can be randomly accessed directly with index expressions (e.g. $C_{ij}$ can be accessed as C[i*N+j]), and these accesses translate into random access requests to memory. 

While RECUMA and REPTILE can lower these recurrences to efficient single-node code, distributed execution introduces a new challenge: data cannot be accessed arbitrarily. Random accesses to memory are either impermissible or impossible when tensors are sharded across memory; there is no guarantee that an arbitrary processor has access to an arbitrary element of a tensor. In both a distributed matrix multiply (over tensors $A$ $B$ and $C$ ) and a distributed triangular solve (over tensors $L$ and $X$ and $B$), a tensor can be either statically distributed over processors, or may stream across processors, such that the same piece of data passes through every processor where it is needed for a computation. 

Regardless if the target architecture is an interconnected lattice of chiplets on the same chip/node, or a grid of CPU processors in a supercomputer, implementing either recurrence in a distributed setting thus requires decisions about how and when data is sharded, streamed, or broadcasted across processing elements. Dependencies in the computations must be respected, and in the cases where multiple pieces of input data are needed for a single computation (e.g. $A_{ik}$ and $B_{kj}$ in matrix multiply), a program must guarantee that all pieces are available at the same processor at the same time, in order to compute with them. In the matrix multiply example, compiling recurrences to distribute processors requires careful handling of:

\begin{enumerate}
    \item \textbf{Communication.}
    The matrix multiply involves three iteration variables $(i,j,k)$, while each tensor uses only two. For instance, $A_{ik}$ lacks $j$, meaning that once $A_{ik}$ is loaded, it can be reused across all $j$ for a fixed $(i,k)$. In a distributed program, this reuse manifests as a communication pattern: $A_{ik}$ may be broadcast or streamed systolically to the processors computing the corresponding $C_{ij}$. Similarly, $B_{kj}$ may be distributed across processors along the $i$ dimension.
    \item \textbf{Dependencies}: Each element $C_{ij}$ can be computed in parallel, but each $C_{ij}$ relies on $A_{ik}$ and $B_{kj}$. Therefore any processor computing $C_{ij}$ at time $t$ must guarantee that $A_{ik}$ and $B_{kj}$ are resident on the same processor at time $t$.
    \item \textbf{Spatial Mapping}: On a 2D grid indexed by $(i,j)$, we can simply assign the computation $C_{ij}$ to processor $(i,j)$. This identity mapping makes $C$ stationary; all computations necessary for computing $C_{ij}$ reside on processor $(i,j)$. Alternatively, it is possible to map $(i,k)$ spatially, making $A$ stationary, or mapping $(j,k)$ spatially, making $B$ stationary
\end{enumerate}

\begin{figure}
\small
\centering
\begin{minipage}{0.42\textwidth}
    \centering
\centering
\footnotesize
\setlength{\abovedisplayskip}{1pt}
\setlength{\belowdisplayskip}{3pt}

{\bf Matrix multiply}\vspace{-2pt}
\[
  C_{ij} = \sum_{k} A_{ik}\,B_{kj}
\]
\vspace{2pt}

{\bf Triangular solve (lower \(L\), forward)}\vspace{-2pt}
\[
  X_{ri} = \frac{1}{L_{ii}}\!\left(B_{ri} - \sum_{j<i} L_{ij}\,X_{rj}\right)
\]
\vspace{2pt}

{\bf Prefix sum}\vspace{-2pt}
\[
  P_i = P_{i-1} + A_i
\]
\vspace{2pt}

{\bf Cholesky (lower \(L\))}\vspace{-1pt}
\[
  L_{ii} = \sqrt{\,A_{ii} - \sum_{k<i} L_{ik}^{2}\,} \;\;
  L_{ij} = \frac{A_{ij} - \sum_{k<j} L_{ik}L_{jk}}{L_{jj}} \; : j<i
\]

\caption{Representative recurrences}
\label{fig:recList}
\end{minipage}
\hfill
\begin{minipage}{0.48\textwidth}
\[
\begin{array}{rlcl} 
    \textsf{ConstInt} & n \\
    \textsf{Tensor} & t \\
    \textsf{IndexVar} & v \\
    \textsf{Index} & i & \bnfdef & v + n \bnfalt n \\
    \textsf{TensorAccess} & ta & \bnfdef & t_{i^+}^{i^*} \\
    \textsf{Expr} & e & \bnfdef & ta \bnfalt  \sum_{v}~e \bnfalt  \textit{const}  \\
     &&& \bnfalt  \sqrt{e} \bnfalt  e + e \bnfalt  e~e \bnfalt  \cdots \\
    \textsf{Recurrence} & r & \bnfdef & ta = e \\
    \textsf{Constraint} & c & \bnfdef & v < i \bnfalt v \leq i \bnfalt v = i \\
    \textsf{Constraints} & cs & \bnfdef & c \bnfalt c, cs \\
    \textsf{ConstrainedRec} & cr & \bnfdef & r : cs \\
    \textsf{Program} & p & \bnfdef & cr \bnfalt cr, p \\
\end{array}
\]
\caption{Recurrence Grammar}
\label{fig:grammar}
\end{minipage}
\end{figure}

Such decisions regarding,dependencies, communication, and spatial mapping result in a large space of distributed programs (including triangular solve and matrix multiply) that compute the same result but with different dataflows. Figure ~\ref{fig:matmulDataFlows} illustrates how algorithm's like Cannons, PUMMA, and weight-stationary systolic multiplication involve different dataflows but all fundamentally conduct a matrix multiply. They all share the same base recurrence $C_{ij} = \sum A_{ik}B_{kj}$, meaning this recurrence in itself is insufficient for uniquely describing each of these matrix multiplication algorithms. 

The same lack of specificity applies to triangular solve and other recurrences. This  warrants a more \emph{specific} form of recurrence equations that define not only the computations but the involved dataflows. These specific recurrences can either serve as a unique input description of the distributed recurrence, or as an intermediate representation that result from feeding the  original recurrences through a lowering algorithm. In the latter case, these more specific, intermediate recurrences can be generated by coupling the original recurrence with additional user-provided scheduling commands that describe data movement.

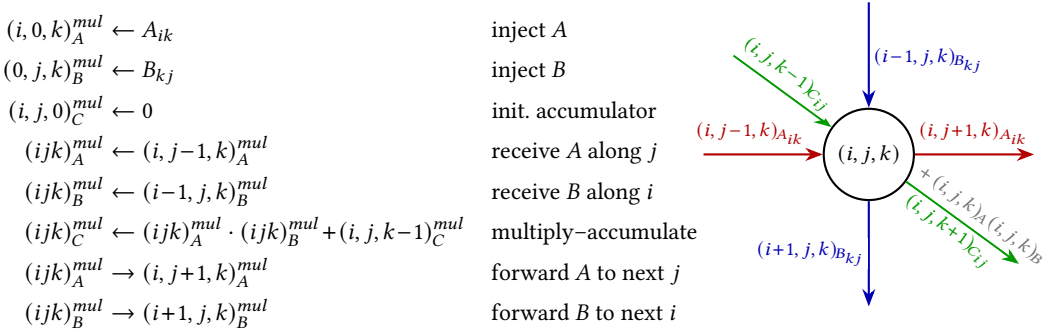
\begin{figure}[t]
\centering
\captionsetup{font=small}
\small

\begin{minipage}[t]{0.62\linewidth}\vspace{0pt}
\[
\setlength{\jot}{2pt}
\begin{aligned}
(i,0,k)_{A}^{mul} &\leftarrow A_{ik} 
    && \text{inject $A$  }\\
(0,j,k)_{B}^{mul} &\leftarrow B_{kj} 
    && \text{inject $B$ } \\
(i,j,0)_{C}^{mul} &\leftarrow 0 
    && \text{init. accumulator} \\
(ijk)_{A}^{mul} &\leftarrow (i,j\!-\!1,k)_{A}^{mul} 
    && \text{receive $A$ along $j$} \\
(ijk)_{B}^{mul} &\leftarrow (i\!-\!1,j,k)_{B}^{mul} 
    && \text{receive $B$ along $i$} \\
(ijk)_{C}^{mul} &\leftarrow 
    (ijk)_{A}^{mul} \cdot (ijk)_{B}^{mul} \!+\! (i,j,k\!-\!1)_{C}^{mul} 
    && \text{multiply--accumulate} \\
(ijk)_{A}^{mul} &\rightarrow (i,j\!+\!1,k)_{A}^{mul} 
    && \text{forward $A$ to next $j$} \\
(ijk)_{B}^{mul} &\rightarrow (i\!+\!1,j,k)_{B}^{mul} 
    && \text{forward $B$ to next $i$}
\end{aligned}
\]
\end{minipage}\hfill
\begin{minipage}[t]{0.34\linewidth}\vspace{0pt}
\centering
\begin{tikzpicture}[
    >=Stealth,
    node font=\footnotesize,
    lab/.style={font=\scriptsize, inner sep=1.2pt, fill=white, align=center},
    Aarr/.style={->,thick,red!70!black},
    Barr/.style={->,thick,blue!70!black},
    Carr/.style={->,thick,green!60!black}
]
\node[draw,circle,minimum size=12mm,thick] (pe) {};
\node at (pe) {$ (i,j,k)$};

\coordinate (A_W) at ($(pe)+(-2.2,0)$);
\coordinate (A_E) at ($(pe)+( 2.2,0)$);
\draw[Aarr] (A_W) -- node[lab,above,sloped,text=red!70!black,pos=.39,yshift=2.2pt] {$(i,j\!-\!1,k)_{\!A_{ik}}$} (pe);
\draw[Aarr] (pe) -- node[lab,above,sloped,text=red!70!black,pos=.49, yshift=2.2pt] {$(i,j\!+\!1,k)_{A_{ik}}$} (A_E);

\coordinate (B_N) at ($(pe)+(0, 2.0)$);
\coordinate (B_S) at ($(pe)+(0,-2.0)$);
\draw[Barr] (B_N) -- node[lab,right,text=blue!70!black,pos=.55] {$(i\!-\!1,j,k)_{\!B_{kj}}$} (pe);
\draw[Barr] (pe) -- node[lab,left,text=blue!70!black,pos=.5] {$(i\!+\!1,j,k)_{\!B_{kj}}$} (B_S);

\coordinate (C_NW) at ($(pe)+(-1.8, 1.3)$);
\coordinate (C_SE) at ($(pe)+( 2.0,-1.44)$);
\draw[Carr] (C_NW) -- node[lab,above,sloped,text=green!60!black,pos=.45] {$(i,j,k\!-\!1)_{\!C_{ij}}$} (pe);
\draw[Carr] (pe) -- node[lab,below,sloped,text=green!60!black,pos=.45] {$(i,j,k\!+\!1)_{\!C_{ij}}$} (C_SE);

\path (pe) -- node[lab,above,sloped,pos=.58,yshift=2pt,text=black!55]
{$+\,(i,j,k)_{\!A}(i,j,k)_{\!B}$} (C_SE);
\end{tikzpicture}
\end{minipage}

\caption{Left: Cyclotron's IR for Cannon's algorithm style GEMM with communication and compute steps annotated. Each line specifies a logical dataflow recurrence over a multi-dimensional iteration space. In this IR, the \emph{loop index tuple} $(i,j,k)$ is the primary entity. The \emph{tensor name} appears as a subscript, indicating which array or value is being updated, and the \emph{iteration space name} appears as a superscript, indicating which recurrence or computation this update belongs to. Loads to memory (e.g., $A_{ik}$) are specified in tensor index notation, which occur at boundaries of the iteration space (e.g., $(i,0,k)$.) Right: Visualization of an internal point of the iteration space}
\label{fig:cannonIR}
\end{figure}

These more specific recurrences describe \emph{iteration spaces with dataflow}, which we refer to as  \emph{dataflow recurrences}. These more specific \emph{dataflow recurrences} specify not only computation but data movement across the iteration space. Figure~\ref{fig:cannonIR} shows the dataflow recurrences that are shared by any form of Cannon's algorithm. In these recurrences, values are owned not by tensors, but by points in the iteration space. Cannon’s algorithm, for example, is describable as a dataflow recurrence, in which $A_{ik}$ and $B_{kj}$ are streamed through nearest neighbors in the 3D iteration space. To illustrate how values of $A_{ik}$ and $B_{kj}$ are owned by different points in the iteration space, we use a notation $(ijk)_A$ to refer to the value of A (e.g., $A_{ik}$) that lives within point $ijk$ in the iteration space. To illustrate how $A_{ik}$  is streamed from the neighboring point (corresponding to the previous $j$), we use the following notation.

$$(ijk)_A = (i,j\!-\!1,k)_A. $$

This dataflow is valid because for a particular $i$ and $k$, $A_{ik}$ is shared/broadcasted across all values of $j$. In the output-stationary Cannon's algorithm, the most basic algorithm for a distributed matrix multiply, we instantiate a 2D grid of processors which are similarly indexed by both $i$ and $j$. This indexing states that processor $(i,j)$ will conduct and own the calculation of output value $C_{ij}$. This dataflow recurrence is amenable to describing the computation in a distributed setting, as it is easy to define a mapping between the 3D iteration space indexed by $(i,j,k)$ and the 2D grid of processors indexed by $(i,j)$. 

Of course, the reference to $(i,j-1,k)_A$ cannot be defined at points where $j=0$, as it breaches the border of the iteration space. Therefore, the value of $(i,0,k)_A$ must be defined as a feeder that reads from tensors. This hence forms a base case for the recurrence, in which all other access to memory otherwise come from neighboring iteration points.  Figure ~\ref{fig:cannonIR} shows the full set of dataflow recurrences needed to implement Cannon's algorithm, with the additional necessary specification that any two of the variables are mapped spatially (usually $i$ and $j$), and the remaining variable is mapped temporally (usually $k$). Similarly, we can decide to index the processing grid by $i$ and $k$, meaning processor $(i,k)$ will own input element $A_{ik}$, such that $A_{ik}$ is \emph{stationary}, and does not move across PEs. When $A_{ik}$ is stationary, output elements $C_{ij}$ and input elements $B_{kj}$ will \emph{stream} through the grid of processing elements. Therefore, any two of the variables from $(i,j,k)$ can be mapped spatially, resulting in different variants of Cannon's algorithm which have the same dataflow among the iteration space but different dataflows between the physical processing elements. 

Alternatively, instead of streaming inputs like $A_{ik}$ across the grid of processors, they can be broadcasted directly to each processor from a common source.  In SUMMA’s algorithm, $A_{ik}$ and $B_{kj}$ are broadcasted globally, in a single step, from a source to all points in the iteration space that require the data. Both algorithms are used in practice, including several hybrid algorithms of the two. In PUMMA’s algorithm, for example, $A_{ik}$ is broadcasted amongst nearest neighbors (as in Cannon’s), but $B_{kj}$ is broadcasted globally, as it is in SUMMA. Triangular solve similarly has a large space of distributed algorithms, as do any recurrence over a multidimensional indexed tensors.

Cyclotron implements these principles in a compiler infrastructure. It allows users to specify recurrences and a set of scheduling primitives describing dataflow, which finally get lowered to a distributed architecture of processing elements. These examples illustrate that even simple recurrences expand into rich spaces of dataflow choices when distributed across processors. Cyclotron’s goal is to make those choices explicit, composable, and compilable.

\section{Overview}
Cyclotron is a compiler that lowers declarative recurrences into programs that execute on distributed architectures. A declarative recurrence is a global equation describing the entire computation, whereas the output programs are local and unique to each processing element (PE) in the distributed system. Each output program consists of compute, send, and receive statements that together implement both computation and communication.

\begin{figure}
    \centering
    \includegraphics[width=0.99\linewidth]{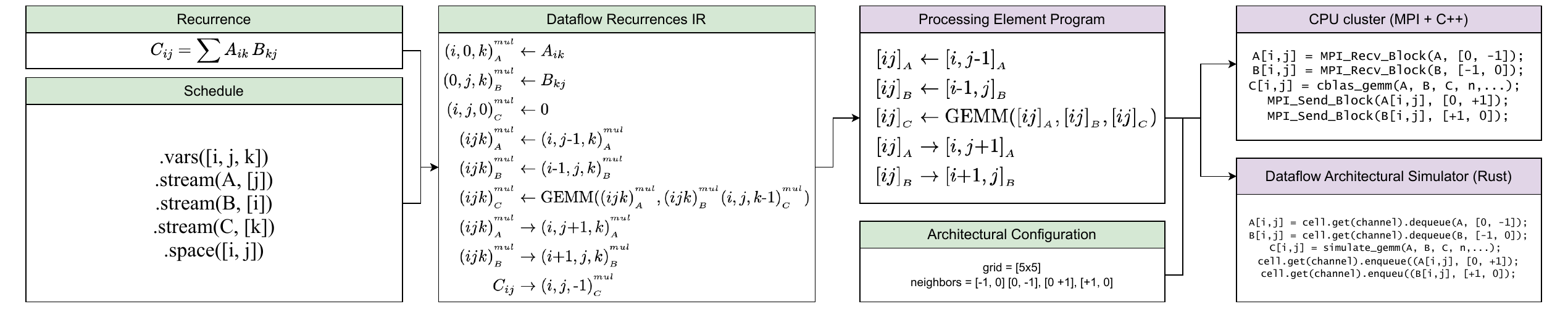}
    \caption{Cyclotron Overview. The system diagram shows the compilation flow for a output-stationary CANNON-style dataflow. By mapping $i$ and $j$ are mapped across space, $Cij$ is stationary, meaning it's locked resident onto each PE and does not move during the computation. $k$ is mapped implicitly across time, such that $A_{ik}$ streams systolically over $j$,  and $B_{kj}$ is stramed over $i$. The IR can be executed on a data-flow simulator of a chiplet style architecture or on an MPI-enabled cluster of processors.}
    \label{fig:overivew}
\end{figure}

Central to Cyclotron is an intermediate representation (IR) of recurrences defined over the iteration space, rather than over tensors as in the input. This intermediate form expresses computation across both spatial and temporal dimensions, yielding an IR of dataflow recurrences that is portable across architectures. Because it abstracts only in terms of send, receive, and compute primitives, the IR can be theoretically lowered to any distributed PE architecture that implements these primitives. Cyclotron demonstrates this by lowering to two distinct targets: (1) a simulator of a dataflow processor composed of interconnected chiplets, and (2) a multinode CPU cluster. Figure ~\ref{fig:overivew} presents an overview of the compiler pipeline, from the declarative recurrence input through the lowering stages to the final backends.

The Cyclotron compiler accepts three user inputs. The first is a set of declarative recurrences defined over tensors. The second is a schedule that specifies which iteration dimensions are mapped to space and time, and how data is streamed across them. Together, these first two inputs describe both the mathematical computation and the dataflow of dependencies and broadcasted values.

Cyclotron compiles the recurrences and schedule into an intermediate representation (IR) of dataflow recurrences that expresses the computation over the program’s iteration space rather than its tensors. This IR is architecture-agnostic. To generate executable programs for a concrete system, Cyclotron requires a third input: a specification of the target architecture and the topology of its processing elements (PEs). The intermediate IR, combined with this architectural specification, is then lowered into a final set of PE-local programs that implement the computation on the target distributed architecture.

Compilation consists of two distinct lowering passes. The first lowering pass takes in the input recurrences and schedule, and then determines what must be computed, sent and received (i.e., communicated) across the program's iteration space. The result of this lowering phase is the aforementioned architectural agnostic IR of \emph{dataflow recurrences}. Here, data accesses come in two flavors: accesses to tensors, and accesses to neighboring points in the iteration space. Tensor reads are denoted by indexed tensor accesses, where the indices are subscripts to the tensor name (e.g., $A_{ik}$). IR statements that read from neighboring points in the iteration spaces are denoted by tensors accesses in which the tensor’s name is a subscript to the current index (e.g., $(ijk)_A$) in the iteration space. This emphasizes that the value is actually owned by a point in the iteration spaces, rather than the tensor, meaning the value can be accessed via communication with the processing element’s neighbors. Users can use scheduling commands that determine whether to \emph{stream}, \emph{broadcast}, or permanently assign (i.e., \emph{prefetch}) data to processing elements. 

The second lowering pass transforms the IR of dataflow recurrences into a set of PE specific programs. Here, users specify which dimensions will be mapped in space (in the diagram, $i$ and $j$ are mapped in space, and $k$ in time). The final IR still uses neighbor access style reads, but in which the time dimension now manifests as a loop over a list of instructions, rather than inside the equations as an additional dimension. The tensor accesses in the output IR can thus be lower dimensional than in the input (e.g., $(ijk)_A$ vs $[ij]_A$, as the time dimension is now removed from the output IR’s tensor accesses. 

Cyclotron generates a separate program for each PE. Having PE-specific programs is necessary, as different PE’s may work on different forms of the computation, may execute different kernels, and have different boundary conditions dictating what gets sent and received. Receives can be directly inferred from data dependencies in the input. Sends are induced via the compiler, which determines where to insert send statements baed on the placement of the receive statements. For any specific receive operation on a PE’s IR, Cyclotron staticaly determines which other PE launches the corresponding send, and then appends a send operation on that other PE’s IR. 

During execution, each processing element interprets each instruction, representing a \emph{megakernel} architecture. One persistent kernel (i.e., function) loops through and executes each instruction. These send, receive, and compute statements are executed in the order they appear on the PE-specific IR. Cyclotron ensures that each instruction in the IR has a corresponding operation in the current backend, whether it be a dataflow simulator that simulates a grid of chiplets, or a cluster of processors. Each target architecture therefore has an ISA, and Cyclotron includes a one-to-one mapping from IR statements to instructions in the architecture’s ISA. This ensures the same executable binary can be run on both targets. This strategically allows for easy porting to different architectures, in which each architecture requires a light runtime (few hundred lines of code) in which each processing element reads in, decodes, and executes each instruction. 

\section{Dataflow Recurrences}

As stated, normal recurrences (of which tensor algebra is a subset) express what must be calculated but do not delineate the data movement and IO necessary to implement the recurrence on a spatial architecture. \emph{Dataflow Recurrences}, which form Cyclotron's intermediate representation, do not have this limitation, as they make communication explicit. Dataflow recurrences, which are defined on top of iteration spaces, have several differences with and features on top of normal recurrences, which are defined on top of tensors. 

\begin{enumerate}
    \item \textbf{Iteration Spaces Owning Data:} Data can be own either by tensors or points in the iteration space, whereas in normal recurrences all data is held within tensors.
    \item \textbf{Explicit Reduction Dimensions:} The dimensionality of all collections is the dimensionality of the enclosing iteration space, not the dimensionality of the underlying tensor.
    \item \textbf{Explicit Communication and Loads into Iteration Space:} Dataflow recurrences define how data is loaded from tensors and onto the iteration space.
    \item \textbf{Multiple Iteration Spaces:} A computation may involve several iteration spaces, which interact through explicit interfaces.
    \item \textbf{Composability with Space-time Mappings} The dataflow recurrences can be lowered to per-processing-element programs with a simple mapping of each variable into space or time.
\end{enumerate}

Consider the canonical matrix multiplication in tensor algebra:

\[
C_{ij} = \sum_{k} A_{ik} \, B_{kj}.
\]

We use this example to illustrate how to transform a \emph{recurrence over tensors} into a \emph{recurrence over iteration spaces}. The equation above is purely \emph{declarative}: it specifies what value each $C_{ij}$ must hold, but not how to schedule the computation or where data must be moved. To derive an executable schedule, we lower this expression into dataflow recurrences, handling all of the aforementioned features. 

\subsection{Iteration Spaces Owning Data}
In normal recurrences, all data is owned by tensors. To illustrate, in matrix multiply, the output values are specified as $C_{ij}$ in which the tensor $C$ is the primary entity, and in which indices appear as a subscript. While dataflow recurrences allow data to be owned by tensors, they importantly also allow data to be owned by the iteration space itself. To denote this, dataflow recurrences use an \emph{indexing first}-notation, in which the loop index tuple $(i,j,k)$ is the primary entity. In this notation, the matrix multiply outputs (including partial outputs) are denoted as $(i,j,k)_C$. A tensor name appears as a subscript, indicating which array or value is being updated, while an iteration-space or recurrence name may appear as a superscript to distinguish different computations. This definition thus places the computation at a precise point $(i,j,k)$ in a 3D iteration space. This notation allows for a natural expression of communication and IO.

\subsection{Explicit Reduction Dimensions}

In the declarative form, the output tensor $C_{ij}$ is indexed by two variables, $i$ and $j$. The reduction index $k$ appears in the expression but does not uniquely identify any output element. In a recurrence over iteration spaces, however, all three iteration variables—$i$, $j$, and $k$—must appear in every reference to $C$, because $C$ is computed inside a triply nested loop. This follows naturally from the dataflow recurrences being defined over the iteration space instead of tensors. We therefore introduce $k$ as an explicit loop index, rewriting the summation as an accumulation over $k$:

\[
(i,j,k)_C \leftarrow (i,j,k\!-\!1)_C + A_{ik} \cdot B_{kj},
\qquad (i,j,0)_C \leftarrow 0.
\]

The reduction manifests as a data dependency between spatially adjacent points along the $k$ axis: $(i,j,k)$ reads from $(i,j,k-1)$. This is a key advantage of iteration-space recurrences: all dependencies map to \emph{neighboring points} in the iteration space, rather than to arbitrary tensor accesses. While explicit tensor accesses are still permissible in Cyclotron, we want all data access to be accesses to neighboring data points as opposed to relatively expensive tensor accesses.


\subsection{Explicit Communication and Loads into Iteration Space} 
Each tensor element $A_{ik}$ is now treated as a value injected
into the iteration space at $j=0$, giving it an explicit spatial
coordinate:

\[
(i,0,k)_A \leftarrow A_{ik}.
\]

This distinguishes the \emph{source of the value} from its
\emph{current location} in the iteration space through which it flows. To make $A_{ik}$ available at every $j$ where $C_{ij}$ is updated,
we introduce a propagation rule along the $j$ axis:

\[
(ijk)_A \leftarrow (i,j\!-\!1,k)_A,
\]

meaning that the copy of $A_{ik}$ needed at $(i,j,k)$ is obtained
by forwarding it from the previous $j$ tile. Syntactically, we notice that in the original recurrence the variable $j$ is missing from the tensor access $A_{ik}$.  In other words, for any particular $i$ and $k$, and for all $j$, the same value of $A_{ik}$ is broadcasted across the $j$ dimension to all points $(i,j,k)$. Therefore the value $A_{ik}$ must be communicated and propagated along the $j$ dimension of the iteration space.

Symmetrically, $B_{kj}$
is injected at $i=0$ and propagated along $i$:

\[
(ijk)_B \leftarrow (i\!-\!1,j,k)_B.
\]

\paragraph{Resulting Space--Time Specification.}
The result is a set of recurrences that jointly specify compute
and communication:

\[
\begin{aligned}
(i,0,k)_A &\leftarrow A_{ik} &\quad&
(ijk)_A \leftarrow (i,j\!-\!1,k)_A \\[2pt]
(0,j,k)_B &\leftarrow B_{kj} &\quad&
(ijk)_B \leftarrow (i\!-\!1,j,k)_B \\[2pt]
(i,j,0)_C &\leftarrow 0 &\quad&
(ijk)_C \leftarrow (i,j,k\!-\!1)_C + (ijk)_A \cdot (ijk)_B.
\end{aligned}
\]

\begin{figure}
    \centering
    \includegraphics[width=0.95\linewidth]{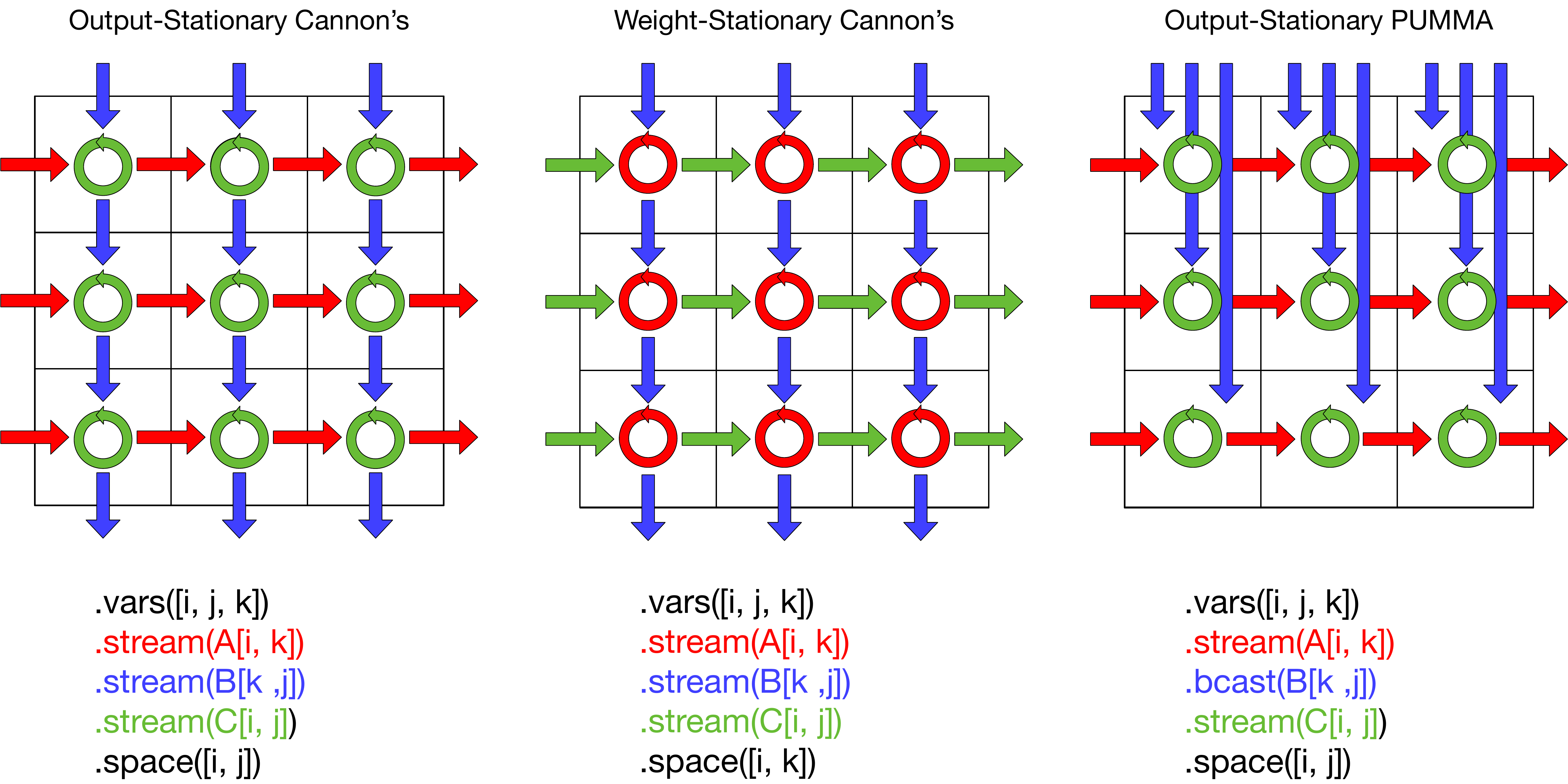}
    \caption{Dataflow of various matrix multiply algorithms. Each algorithm implements the recurrences $C_{ij} = A_{ik}B_{kj}$, but with different dataflow and communication. Left, output stationary Cannon's algorithm, in which $A$ and $B$ both stream systolically across the grid of processing elements. Middle: Weight (e.g., $A_{ik}$ stationary Cannon's algorithm, in which $B$ and $C$ are streamed in. Right: Output stationary PUMMA algorithm. This algorithm broadcasts $B$ down the i dimension, and systolically streams $A$ in the $j$ direction. }
    \label{fig:matmulDataFlows}
\end{figure}

Notably all communication is with nearest neighbors, with any offsets to index variables being $\!-\!1$. This formulation makes the dataflow explicit. Output-stationary Cannon's algorithm implements these dataflow recurrences such that $i$ and $j$ are mapped across space, and $k$ across time. This is visualized by Figure~\ref{fig:matmulDataFlows} . Here, $A$ tiles stream
rightward across $j$, $B$ tiles stream downward across $i$, and
each $(i,j)$ processing element accumulates $C_{ij}$ over $k$. The final write to the result tensor $C$ can then be optionally specified with the recurrence $C_{ij} \leftarrow (i,j,k_{end})_C$, which represents exporting data from the iteration space to memory.

\subsection{Multiple Iteration Spaces}

Many computations require more than one iteration space to express their structure and data dependencies. In Cyclotron, each distinct operation in the recurrence’s abstract syntax tree—such as \texttt{GEMM}, \texttt{TRSM}, or \texttt{SQRT}—is assigned its own iteration space, defining when and where that operation occurs in space–time. A simple matrix multiplication involves only one such operation and therefore a single iteration space. In contrast, composite computations like block triangular solve (\texttt{TRSM}) involve multiple sub-operations, each with its own spatial and temporal pattern.

\begin{minipage}[t]{0.45\linewidth}
\centering
\[
X_i \;=\;
\operatorname{TRSM}\!\Big(
L_{ii},\;
B_i \;-\;
\sum_{j < i}
    \operatorname{GEMM}\!\big(L_{ij},\, X_j\big)
\Big)
\]
\end{minipage}%
\hfill
\begin{minipage}[t]{0.45\linewidth}
\centering
\begin{lstlisting}[language={},mathescape=true,basicstyle=\ttfamily\small,frame=single]
X_i = B_i
for j < i:
    X_i -= GEMM(L_ij, X_j)
X_i = TRSM(L_ii, X_i)
\end{lstlisting}
\end{minipage}

\begin{figure}
    \centering
    \includegraphics[width=0.4\linewidth]{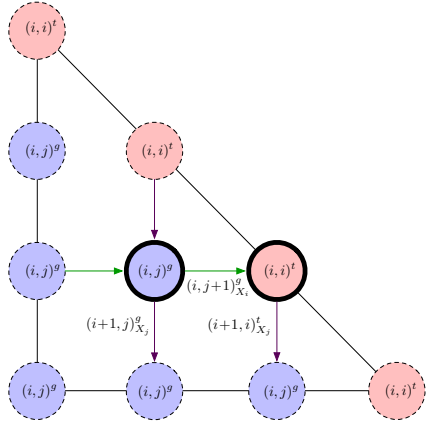}
    \caption{Interlinked iteration spaces in the triangular solve, with sends labeled for the two bold nodes. The diagonal spa ce denoted by $(i,i)$ points conduct local triangular solves. The lower triangular space denoted by $(i,j)$ points conduct local GEMMs. The two spaces interact at the boundary. The two bold points(1 gemm point, and 1 trisolve point) are also illustrated in Figure ~\ref{fig:trisSolveDataflow}}
    \label{fig:triSpace}
\end{figure}
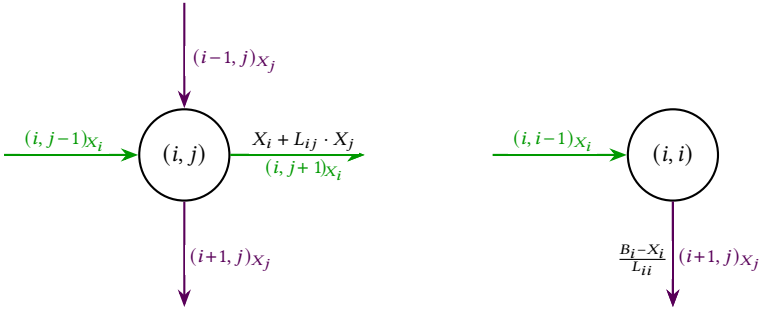
\begin{figure}
    \centering

\begin{tikzpicture}[
  >=Stealth,
  node font=\small,
  lab/.style={font=\scriptsize, inner sep=1.2pt, fill=white, align=center},
  Xb/.style={->,thick,violet!70!black},   
  Xa/.style={->,thick,green!60!black}     
]
\node[draw,circle,minimum size=12mm,thick] (pe) {};
\node at (pe) {$ (i,j)$};

\coordinate (N) at ($(pe)+(0,2.0)$);
\coordinate (S) at ($(pe)+(0,-2.0)$);
\coordinate (W) at ($(pe)+(-2.4,0)$);
\coordinate (E) at ($(pe)+(2.4,0)$);

\draw[Xb] (N) -- node[lab,right,text=violet!70!black,pos=.55] {$\,(i\!-\!1,j)_{X_{j}}$} (pe);
\draw[Xb] (pe) -- node[lab,right,text=violet!70!black,pos=.55] {$(i\!+\!1,j)_{X_{j}}$} (S);

\draw[Xa] (W) -- node[lab,above,sloped,text=green!60!black,pos=.45] {$(i,j\!-\!1)_{\!X_{i}}$} (pe);
\draw[Xa] (pe) -- node[lab,below,sloped,text=green!60!black,pos=.55] {$(i,j\!+1\!)_{\!X_{i}}$} (E);
\draw[Xa] (pe) -- node[lab,above,sloped,text=black!60!black,pos=.55] {$X_{i}+L_{ij}\cdot X_{j}$} (E);

\end{tikzpicture}
\hspace{1.5cm}
\begin{tikzpicture}[
  >=Stealth,
  node font=\small,
  lab/.style={font=\scriptsize, inner sep=1.2pt, fill=white, align=center},
  Xb/.style={->,thick,violet!70!black},
  Xa/.style={->,thick,green!60!black},
  Bv/.style={->,thick,blue!70!black}
]
\node[draw,circle,minimum size=12mm,thick] (pe2) {};
\node at (pe2) {$(i,i)$};

\coordinate (N2) at ($(pe2)+(0,2.0)$);
\coordinate (S2) at ($(pe2)+(0,-2.0)$);
\coordinate (W2) at ($(pe2)+(-2.4,0)$);

\draw[Xa] (W2) -- node[lab,above,sloped,text=green!60!black,pos=.45] {$(i,i\!-\!1)_{X_{i}}$} (pe2);

\draw[Xb] (pe2) -- node[lab,right,text=violet!70!black,pos=.55] {$(i\!+\!1,j)_{X_{j}}$} (S2);
\draw[Xb] (pe2) -- node[lab,left,text=black!70!black,pos=.55] { $\frac{B_{i} - X_{i}}{L_{ii}}$} (S2);

\end{tikzpicture}
    \caption{Dataflow of each operation in the triangular solve.  Left: dataflow of the GEMM within the triangular solve. Right: dataflow of the smaller triangular solve within the triangular solve. The triangular solve node ingests outputs of the GEMMS, and then produces values that seed future GEMMs}
    \label{fig:trisSolveDataflow}
\end{figure}

Figures ~\ref{fig:triSpace} and \ref{fig:trisSolveDataflow}
visualize the dataflow of a tiled lower-triangular solve. Figure ~\ref{fig:deps-tri-solve} shows the corresponding dataflow recurrences. The computation decomposes into two coupled spaces. There's a \emph{triangular space} performing streamed GEMM updates, and a \emph{diagonal space} performing smaller, recursive TRSM solves. 

Each space carries its own recurrences, loads, and local communications. The triangular space accumulates updates along rows and columns, while the diagonal space periodically consumes a boundary of the triangular space to perform a smaller triangular solve. Once a diagonal solve completes, its result is broadcast back into the triangular space to seed the next wave of updates. This interaction is captured by explicit \emph{interface maps} between the iteration spaces, denoted $\Phi_1$ and $\Phi_2$, which define how coordinates and data are translated between their domains.



\begin{figure}[t]
\centering
\footnotesize
\begin{minipage}[t]{0.98\linewidth}
\begingroup
\setlength{\jot}{2pt}

\[
\textbf{Spaces:}\quad 
\mathcal D_{\mathrm{tri}}=\{(i,j)\mid 0\le j\le i<n\},\qquad
\mathcal D_{\mathrm{diag}}=\{(ii,ii)\mid 0\le ii<n\}.
\]

\begin{center} \textbf{Initialization / loads:}\end{center}
\[
\begin{aligned}
(i,j)^{\mathrm{tri}}_{L} &\leftarrow L_{ij} 
    && \text{stream tiles of }L \\
(ii,ii)^{\mathrm{diag}}_{L} &\leftarrow L_{ii,ii},\quad
 (ii,ii)^{\mathrm{diag}}_{B} \leftarrow B_{ii}
    && \text{diag loads}
\end{aligned}
\]

\begin{center}\textbf{Boundary handshakes (tri $\leftrightarrow$ diag):} \end{center}
\[
\begin{aligned}
(ii,ii)^{\mathrm{diag}}_{X_{a}} &\leftarrow (ii,ii\!-\!1)^{\mathrm{tri}}_{X_a}
    && \text{TRSM reads end-of-row from tri} \\
(i,j_e)^{\mathrm{tri}}_{X_b} &\leftarrow (i\!-\!1,j)^{\mathrm{diag}}_{X} 
    && \text{optional: seed broadcast at } j_e
\end{aligned}
\]

\begin{center}\textbf{Intra-space data movement:}\end{center}
\[
\begin{aligned}
(i,j)^{\mathrm{tri}}_{X_b} &\leftarrow (i\!-\!1,j)^{\mathrm{tri}}_{X_b}
    && \text{vertical broadcast in column } j \\
(i,j)^{\mathrm{tri}}_{X_a} &\rightarrow (i,j\!+\!1)^{\mathrm{tri}}_{X_a}
    && \text{forward accumulator along } j
\end{aligned}
\]


\begin{center}\textbf{Computation (tri space: GEMM)}\end{center}
\[
\begin{aligned}
(i,j)^{\mathrm{tri}}_{X_a} 
    &\leftarrow \mathrm{GEMM}\;\!\Big((i,j)^{\mathrm{tri}}_{L},\ (i,j)^{\mathrm{tri}}_{X_b}\Big)
    && \text{for } i> j,\
\end{aligned}
\]

\begin{center}\textbf{Computation (diag space: TRSM):}\end{center}
\[
(ii,ii)^{\mathrm{diag}}_{X} 
    \leftarrow \mathrm{TRSM}\;\!\Big((ii,ii)^{\mathrm{diag}}_{L},\ (ii,ii)^{diag}_{B} \!-\!(ii,ii)^{\mathrm{diag}}_{X_{a}}\Big)
\]

\begin{center}\textbf{Publish (optional):}\end{center}
\[
X_{ii} \leftarrow (ii,ii)^{\mathrm{diag}}_{X}.
\]

\endgroup
\end{minipage}
\vspace{-0.5ex}
\caption{Dataflow recurrences i.e. IR for block lower-triangular solve $LX=B$ with two iteration spaces:
triangular space (streamed $\mathrm{GEMM}$) and diagonal space (TRSM).}
\label{fig:deps-tri-solve}
\end{figure}

\begin{center}\textbf{Interface maps:}\end{center}
\[
\begin{aligned}
\Phi_{1} :\;& \mathcal D_{\mathrm{diag}} \;\rightarrow\; \mathcal D_{\mathrm{tri}}^{\rightarrow},
&\quad& \Phi_{1}(ii,ii) = (i{=}ii,\ j{=}j_e{+}1), \\[4pt]
\Phi_{2} :\;& \mathrm{diag}(\mathcal D_{\mathrm{tri}}) \;\rightarrow\; \mathcal D_{\mathrm{tri}}^{\searrow},
&\quad& \Phi_{2}(j,j) = (j{+}1,\ j).
\end{aligned}
\]

The resulting dataflow recurrences (Figure~\ref{fig:deps-tri-solve}) show two iteration spaces operating concurrently but exchanging data through explicit, one-to-one interfaces. Each space defines its own local schedule, while the interface maps $\Phi_1$ and $\Phi_2$ synchronize their boundaries in both time and index coordinates. Together they form a unified spacetime representation of the block triangular solve: every transfer, update, and synchronization is made explicit in the IR, allowing Cyclotron to reason about both computation and communication at the same level of abstraction.

 Figure ~\ref{fig:deps-tri-solve} shows the dataflow recurrences for these two iteration spaces, one for the GEMM part of the computation, and one for the  triangular solve (TRSM) part of the computation. Each iteration space has it's own dataflow, with the internal GEMM computation receiving data other GEMM computations. The boundary GEMMS require special handling; initial GEMMS receive their data from TRSM, and terminal GEMMs send their partial outputs to TRSM.  This is made explicit by the \textbf{boundary-handshakes } in Figure ~\ref{fig:deps-tri-solve},  in which the two spaces exchange values via the interface maps. The exchange can also bee seen in the interface between the diagonal $(i,i)$ nodes and triangular  $(i,j)$ nodes in  Figure ~\ref{fig:triSpace} , showing the full triangular iteration space. 

\begin{figure}[t]
\centering
    \includegraphics[width=0.9\linewidth]{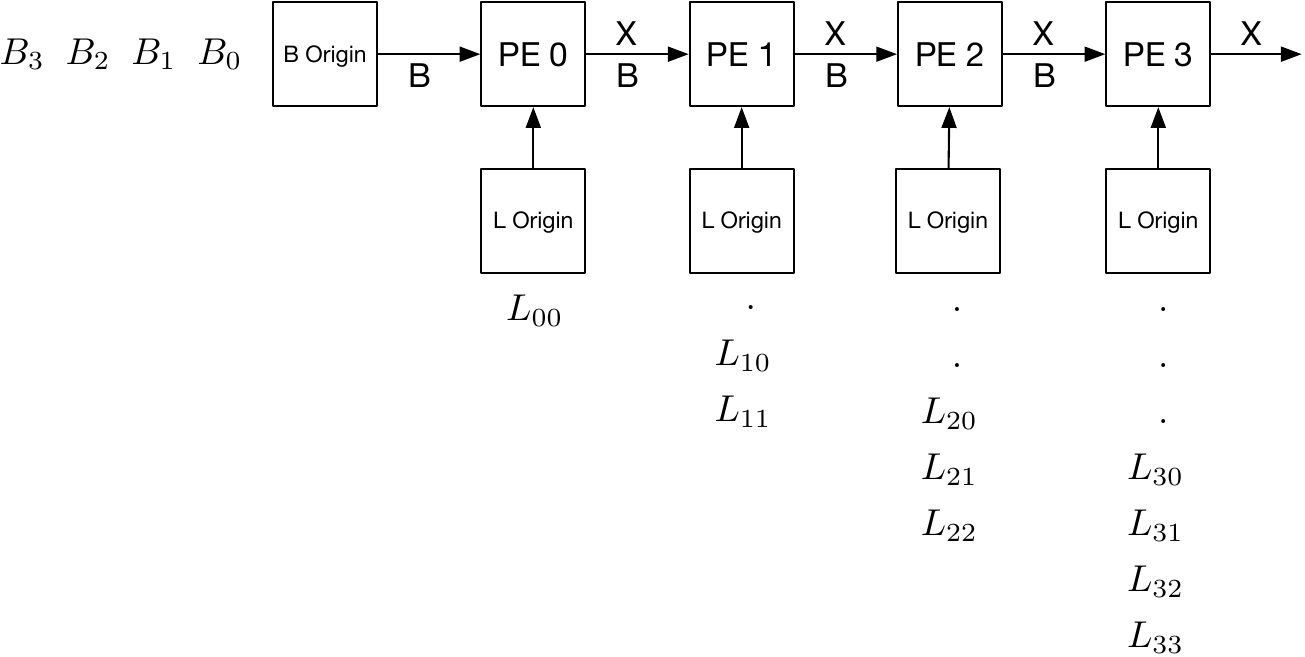}
    \caption{Distributed triangular solve across processing elements.Each PE reads in one block row of $L$ and the corresponding partition of $B$. Diagonal blocks $L_{ii}$ perform local solves, while off-diagonal $L_{ij}$ blocks perform matrix-vector multiplies and propagate partial results $x_j$ to subsequent PEs. Arrows indicate the flow of $x$ and $b$ values and the sequential dependencies implied by the recurrence.}
    \label{fig:triSystolicBase}
\end{figure}

\section{Scheduling}

Cyclotron provides a few scheduling commands to the user: \emph{stream}, \emph{prefetch}, and \emph{broadcast}. We show how all three can be used in the context of a tiled triangular solve, resulting in systolic arrays with different performance characteristics. Here, the equation for the tiled triangular solve is: 

\[
x_i \;=\;
\operatorname{TRISOLVE}\!\Big(
L_{ii},\;
b_i \;-\;
\sum_{j < i} \operatorname{GEMV}\!\big(L_{ij},\, x_j\big)
\Big)
\]

The triangular solve can be scheduled in different ways, such that the user can configure where
data for $B$ or $L$ is stationary or streamed. If $B$ is \emph{streamed}, the $i$th PE consumes the $ith$ portion of $B$ as it arrives, performing local triangular solves using its resident $L_{ii}$ block before forwarding the newly computed $X_i$ downstream. In this configuration (Figure ~\ref{fig:triSystolicBase}) $L$ is also streamed, and each PE consumes portions of $L_{ij}$ as they arrive, performing a matrix vector multiply between tile $L_{ij}$ and vector slice $x_j$ and then forwarding $x_j$ downstream to other PEs where it will be needed. This pattern is expressed in Cyclotron using the \texttt{stream} directive:
\begin{lstlisting}[language={},basicstyle=\ttfamily\small,mathescape=true]
stream(B, [i])
\end{lstlisting}

Figure ~\ref{fig:triSystolicBase} shows the 1D PE grid implementing a triangular solve, where $L$, $B$ and $X$ are all streamed. Alternatively, as $i$ is mapped across space, we can presave all input $B_i$s across PEs in the $i$ dimension, such that $B$ becomes \emph{stationary}: each PE holds its $B_i$ segment locally and consumes incoming blocks of $L_{ij}$ as they arrive. This pattern becomes useful when reusing $B$ across different $L$ matrices, and can be expressed with the \texttt{prefetch} directive:
\begin{lstlisting}[language={},basicstyle=\ttfamily\small,mathescape=true]
prefetch(B, [i])
\end{lstlisting}

\begin{figure}[t]
\centering
\small
\begin{minipage}[t]{0.4\linewidth}
    \centering
    \includegraphics[width=\linewidth]{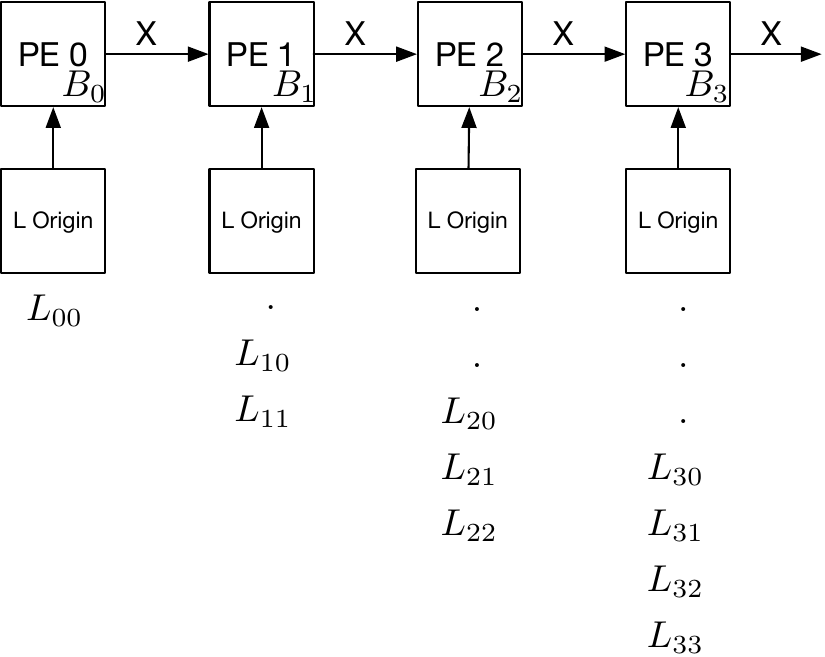}
    \caption{prefetch $B$}
    \label{fig:triPreload}
\end{minipage}
\hfill
\begin{minipage}[t]{0.45\linewidth}
    \centering
    \includegraphics[width=\linewidth]{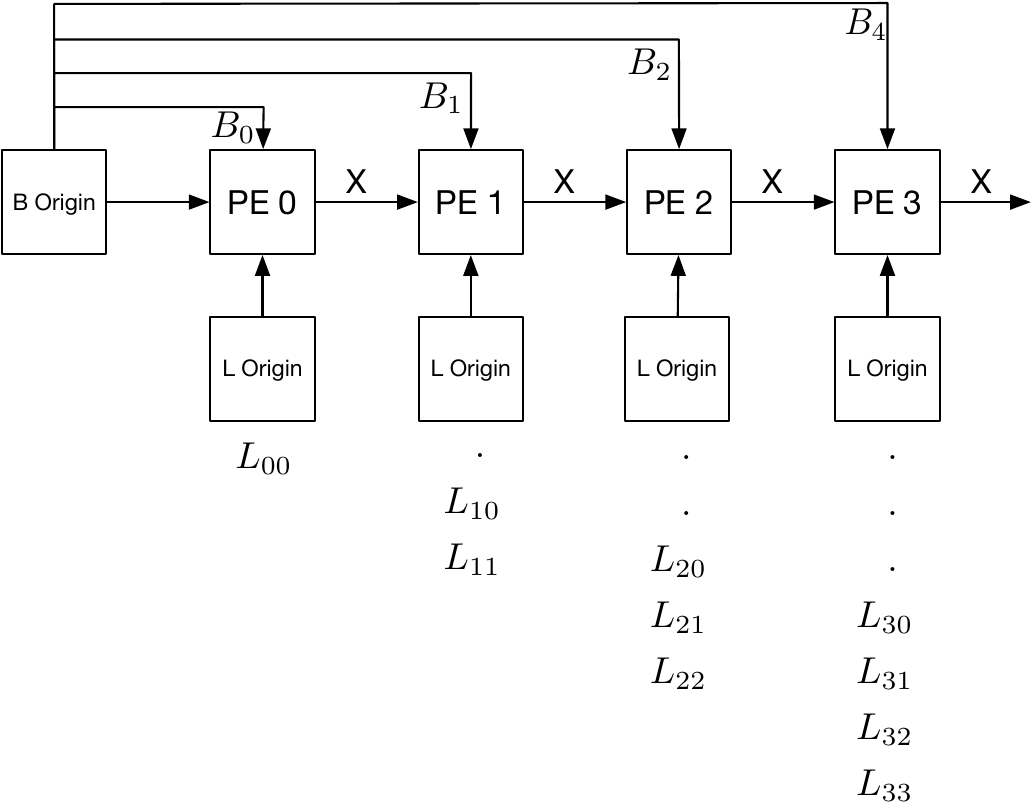}
    \caption{broadcast $B$}
    \label{fig:triBcast}
\end{minipage}
\caption{Two systolic arrays of a blocked triangular solve. Left: B is stationary, and L is streamed. Right, L is streamed and B is broadcasted.In both cases, $L_{ij}$ is skewed by it's column offset $j$ in order to ensure it arrives at PE $i$ at the same time as $X_j$}
\end{figure}

Figure ~\ref{fig:triPreload} shows a 1D PE grid created by scheduling command prefetch($B$, [$i$]), meaning $B_i$ is now stationary, as opposed to streamed into the grid as it was in Figure ~\ref{fig:triSystolicBase}/
 Finally, if there is a broadcast network for $B$, such that every PE has a connection to the source of $B$ (i.e., the \emph{feeder}), then each PE can receive $B$ directly from the feeder. This is analogous to the broadcasts seen in the PUMMA (and SUMMA) algorithm; the matrix is broadcasted directly from the feeder, rather then streamed through PEs. Figure ~\ref{fig:triBcast} shows a 1D PE grid in which $B$ is broadcasted individually to all processors. This can be accomplished with the \texttt{broadcast} directive: 
\begin{lstlisting}[language={},basicstyle=\ttfamily\small,mathescape=true]
broadcast(B, [i])
\end{lstlisting}

These three schedules are simply different implementations of the same base recurrence for a tiled triangular solve. In practice, the optimal choice between them depends on the existence of a broadcast network, whether $B$ or $L$ can be reused, and the bandwidth of all of the links.

\section{Space-time Mappings of Dataflow Recurrences}

The recurrences above define dataflow over a three-dimensional \emph{logical} iteration space, not over any concrete hardware. Given, for example, a two-dimensional array of physical processors, any two iteration dimensions may be mapped to \emph{space}, with the remaining one implicitly mapped to \emph{time}. Intuitively, spatial dimensions correspond to parallelism across processors, while temporal dimensions correspond to values that are accumulated locally over time. Choosing which dimensions to assign to space versus time is a user-specified scheduling decision, and different mappings are advantageous in different settings.

If we choose to map $i$ and $j$ in space, then $k$ is implicitly mapped to time. Communication along the $i$ or $j$ axes therefore manifests as physical communication across the processor array, whereas communication along the $k$ axis simply corresponds to reading a value from a previous time step. Folding $k$ into time yields the following program. This program represents the dataflow of all PEs, using constant indices (e.g. $[i,0]$ ) to indicate special handling of boundary PEs:

\[
\begin{aligned}
[i,j]_C \leftarrow 0 \\[6pt]
\textbf{for } k\! =\! 1,2,\dots \\
[i,0]_A &\leftarrow A_{ik} \\
[0,j]_B & \leftarrow B_{kj} \\
\quad [i,j]_A &\leftarrow [i,j\!-\!1]_A \\[2pt]
\quad [i,j]_B &\leftarrow [i\!-\!1,j]_B \\[2pt]
\quad [i,j]_C &\leftarrow [i,j]_C + [i,j]_A \cdot [i,j]_B \\[2pt]
\quad [i,j]_A &\rightarrow [i,j\!+\!1]_A \\[2pt]
\quad [i,j]_B &\rightarrow [i\!+\!1,j]_B \\[2pt]
\end{aligned}
\]


In this schedule, $A$ and $B$ stream across hardware links, while $C$ remains \emph{stationary}. However, the roles can be inverted. In scenarios where $A$ is much larger than $B$ and $C$—as in neural networks where $A$ contains the weights— it is preferable to keep $A$ resident and instead move the activations. This is achieved by mapping $i$ and $k$ in space, and $j$ in time, yielding the well-known \emph{weight-stationary} schedule from neural network accelerators, where partial outputs of $C$ traverse the systolic array at each timestep:



\[
\begin{aligned}
[i,k]_W &\leftarrow A_{ik} \\[4pt]
\textbf{for } j = 1,2,\dots \\[-2pt]
\quad [0,k]_B &\leftarrow B_{kj} 
\;\qquad\qquad\qquad\qquad\qquad\qquad\text{(inject at } i{=}0\text{ i.e., north boundary)}\\
\quad [i,0]_C &\leftarrow 0 \qquad\qquad\qquad \quad\quad\quad\quad\quad\quad\quad\;\text{(init. accumulator  at west boundary)}\\
\quad [i,k]_B &\leftarrow [i\!-\!1,k]_B \qquad\qquad\qquad\qquad\qquad\text{(receive $B$ from north, non-boundary case)}\\[4pt]
\quad [i,k]_C &\leftarrow [i,k\!-\!1]_C \;+\; [i,k]_W \cdot [i,k]_B \;\qquad\text{(receive $C$ and perform GEMM)}\\[4pt]
\quad [i,k]_C &\rightarrow [i,k\!+\!1]_C  \quad\quad\quad\quad\quad\quad\quad \quad\quad\quad\text{(forward $C$ east along $k$, non-boundary case)}\\[4pt]
\quad [i,k]_B &\rightarrow [i\!+\!1,k]_C \quad\quad\quad\quad\quad\quad\quad \quad\quad\quad \text{(forward $B$)}
\end{aligned}
\]


Lowering from the dataflow recurrences to per-PE programs proceeds by a simple \emph{pattern matching} of each point in the processor grid against the iteration-space templates defined by the recurrences. Each recurrence specifies a dataflow rule over a 3D iteration space $(i,j,k)$; during lowering, Cyclotron instantiates these rules for each processor coordinate, selecting only those whose index patterns mostly closely match the processor’s spatial location and boundary conditions. 

In the \emph{weight-stationary} variant of Cannon's algorithm, $(i,k)$ are mapped spatially and $j$ advances in time. A processor at cell $(i,k\!=\!0)$ matches the rule $(i,j,k\!=\!0)_{C}^{mul} \leftarrow 0$, which initializes the local partial sum $C_{ij}$ along the west boundary. Likewise, processors at $(i\!=\!0,k)$ on the northern boundary bind to the inject rule $(i\!=\!0,j,k)_{B}^{mul} \leftarrow B_{kj}$ rather than the streaming form $(ijk)_{B}^{mul} \leftarrow (i{-}1,j,k)_{B}^{mul}$.


This \emph{template-matching} process produces one localized program per cell type (corner, edge, or interior). When the temporal dimension $j$ is folded into an outer loop, the resulting programs correspond to the per-PE kernels shown in Figure~\ref{fig:perPE}.

Conceptually, this lowering stage replaces dataflow recurrences with concrete per-processor programs obtained by instantiating only the rules that match each cell’s coordinates.

\hfuzz=20pt      
\sloppy      
\begin{figure}[t]  
\scriptsize
\raggedright
\setlength{\fboxsep}{2pt}
\setlength{\fboxrule}{0.4pt}

\firstcell\hspace{0.1em}\secondcell\hspace{0.1em}\thirdcell\\[0.3em]
\fourthcell\hspace{0.1em}\fifthcell\hspace{0.1em}\sixthcell\\[0.3em]
\seventhcell\hspace{0.1em}\eigthcell\hspace{0.1em}\ninthcell
\caption{Localized per-cell programs for a weight-stationary Cannon's , showing both the internal PE and all eight boundary conditions. Each boxed cell represents the computation executed by a processing element (PE) at spatial coordinates $(i,k)$ over time coordinate $k$ . \textcolor{Ared}{Red} statements denote loads of $A_{ik}$ into the edge of the systolic array, whereas \textcolor{Bblue}{blue} statements denote loads of $B_{kj}$ into the systolic array, and \textcolor{Cgreen}{green} statements denote write of $C_{ij}$ among the output edge. For a matrix multiply, cyclotron will emit these 9 types of programs, with programs duplicated across PEs that play the same role (e.g., all internal cells will have the same program denoted by Cell $(i,j)$, as they do not require I/O with memory.}
\label{fig:perPE}
\end{figure}

\section{Timing}

For any systolic execution, correctness requires that all operands of each recurrence---such as $A_{ik}$ and $B_{kj}$ in matrix multiplication---arrive at the same processing element (PE) and at the same time step. Misalignment would cause a stall or the use of stale data. Intuitively, the schedule defines a space--time wavefront that sweeps through the array: each PE on that wavefront receives exactly the operands it needs from orthogonal directions at the same clock tick.

The relative timing between operands arises from a \emph{skew} in the schedule.When one loop index is mapped to time, values associated with different spatial coordinates are emitted at staggered offsets so that their trajectories through
space intersect at the correct moment. In other words, temporal skew encodes how far a value must travel before being consumed. In matrix multiplication,
for instance, $A_{ik}$ and $B_{kj}$ are launched at different phases: $A_{ik}$ is delayed by its horizontal offset $i$, while $B_{kj}$ is delayed by its vertical offset $j$. These coordinated delays ensure that both operands reach
PE$(i,j)$ simultaneously, inducing the characteristic diagonal wavefront of systolic execution.

This alignment can be expressed geometrically.  
Mapping any two indices to space and the remaining one to time induces a global execution time 
\[
t = i + j + k,
\]
which represents the \emph{Manhattan distance} from the origin of the iteration space to point $(i,j,k)$ under unit communication latency.  
Each operand’s coordinates determine its emission time and propagation delay so that all dependencies for $(i,j,k)$ meet at PE$(i,j)$ precisely at $t=i{+}j{+}k$.  
Cyclotron verifies this invariant algebraically when generating distributed schedules.

The same reasoning extends to dependent recurrences such as triangular solve.  
In Figure~\ref{fig:triSystolicBase}, $i$ is mapped in space, $j$ in time, and tensors $L$, $X$, and $B$ are streamed into the array.  
Operand $L_{ij}$ originates with an offset of $i$ and advances for $j$ steps before reaching PE~$i$, arriving at time $i{+}j$.  
Meanwhile, $X_j$ is produced at time $2j$ (after $j$ internal reductions) and then traverses $i{-}j$ PEs to reach PE~$i$ at the same time $i{+}j$.  
Thus both $L_{ij}$ and $X_j$ arrive concurrently to perform $L_{ij}\,X_j$.  

This unified geometric interpretation---where communication latency corresponds to Manhattan distance in the iteration space---ensures that all operands intersect at the correct PE and time step.

\newcommand{\I}{\vec{i}}  
\newcommand{\ridx}{r}
\newcommand{\X}[1]{X^{(#1)}}
\newcommand{\g}[1]{g_{#1}}
\newcommand{\op}{\mathbin{\oplus}}
\newcommand{\Ff}{F}
\newcommand{\stream}{\mathsf{stream}}
\newcommand{\upd}[3]{#1[#2 \mapsto #3]}
\newcommand{\Id}{\mathsf{id}}

\section{Lowering Recurrences to Dataflow Recurrences}
The input recurrences are lowered to dataflow recurrences via a set of rewrite rules, which dictate how to transform accesses to indexed tensors into the streaming dataflow. In the matrix multiply example, our rewrite rules  rewrite the tensor access $A_{ik}$ into streaming dataflow recurrences, which include $(i,j,k)_{A}\leftarrow(i,j-1,k)$ and the boundary case $(i,j,k)_{A} \leftarrow A_{ik}$

Consider output indices $\I$, reduction index $\ridx$, output tensor $C$, and input operands $\X{t}$ with index maps $\g{t}(\I,\ridx)$ (i.e., it maps $(ijk)_A$ to $ik$)
Each operand chooses a stream axis $s_t \in \I$. Finally, let  $r_{\mathrm{end}}(\I)$ be the exclusive upper bound of reduction variable $r$, and let $r_{\mathrm{final}}(\I)$ be it's inclusive upper bound.

\newcommand{\R}{\vec{r}}                      
\newcommand{\rfin}{\R_{\mathrm{final}}(\I)}   
\newcommand{\rend}{\R_{\mathrm{end}}(\I)}   
\newcommand{\indices}[1]{\mathrm{indices}(#1)}
\newcommand{\cE}{\mathcal{E}}                 
\newcommand{\tilC}{\widetilde{C}}             
\newcommand{\Dom}{\mathcal{D}}                
\newcommand{\Ready}{\mathsf{Ready}}           
\newcommand{\Buf}{\mathsf{Buf}}               
\newcommand{\phiab}{\varphi_{a\to b}}         
\[
C_{o(\I)}
\;=\;
\cE_{\I}\!\Bigg[
  \sum_{\R }
    F\!\big(
       X^{(1)}_{\,g_1(\I,\R)},
       \ldots,
       X^{(m)}_{\,g_m(\I,\R)},
       \; C_{\,g_C(\I,\R)} \;\text{(optional)}
     \big)
\Bigg].
\]

\begin{mathpar}

\inferrule*[right=ChooseStream]
{ s_t \in \I \qquad s_t \notin \indices{\X{t}} }
{ \stream(\X{t}, s_t) }

\inferrule*[right=LowerSum]
{ }
{ (\I,\mathbf{0})_{\tilC} \leftarrow 0 \qquad
  (\I,\R)_{\tilC} \leftarrow (\I,\R\!-\!1)_{\tilC}
    \;+\;
    F\!\big( \X{1}_{\g{1}(\I,\R)}, \dots, \X{m}_{\g{m}(\I,\R)},\; C_{h(\I,\R)}^{\text{(opt.)}} \big)
  \\
  \text{(If }\R=\emptyset\text{, interpret }(\I,\mathbf{0})_{\tilC} := F(\dots)\text{.)}
}

\inferrule*[right=Inject]
{ \stream(\X{t}, s_t) \qquad \mathbf{p}_t = \g{t}(\I,\R) }
{ (\upd{\I}{s_t}{0},\R)_{\X{t}} \leftarrow \X{t}_{\mathbf{p}_t} }

\inferrule*[right=Propagate]
{ \stream(\X{t}, s_t) }
{ (\I,\R)_{\X{t}} \leftarrow (\upd{\I}{s_t}{s_t-1},\R)_{\X{t}} }

\inferrule*[right=Send]      
{ \stream(\X{t}, s_t) }
{ (\I,\R)_{\X{t}} \rightarrow (\upd{\I}{s_t}{s_t+1},\R)_{\X{t}} }

\end{mathpar}




\begin{mathpar}
\inferrule*[right=OuterSend]
{ \cE_{\I}[\cdot] \neq [\cdot] \;\;\; \text{i.e. summation has outer expression} }
{ (\I,\rfin)_{\tilC}^{inner} \;\rightarrow\; (\I, \rend)^{outer}_{\tilC} \quad
  (\I, \rend)_{\tilC}^{outer} \leftarrow(\I, \rfin)^{inner}_{\tilC} \\(\I, \rend)_{C}^{outer}  \leftarrow  \cE_{\I}\!\big[\,(\I, \rend)^{outer}_{\tilC}\,\big]  } 
\end{mathpar}



\begin{mathpar}
\inferrule*[right=InnerSend]
{ \text{$C$ used in $F$ and} \;\; s_t \in \I \qquad \text{s.t.} \;\; r_{\mathrm{end}}(\I) = s_t }
{ 
   (\I,\, r_{\mathrm{final}}(\I))_{C}^{inner}
   \;\leftarrow\;
   \big(\,\upd{\I}{s_t}{s_t\!-\!1},\, r\,\big)_{C}^{outer} }
\end{mathpar}

 The transformation proceeds by a sequence of syntactic rewrites, shown formally in the inference rules. The key idea is to replace the implicit summation with a running accumulator over the reduction index, and to materialize the flow of each operand as a stream through the output iteration space. 

The first five rules are the \emph{core} rules that define how to handle lowering a \emph{stream()} command within the single iteration space. These rules (which do not include the inner send and outer) are shared between recurrences with and and without dependencies. They define how data moves within a single iteration space. The outer/inner send rules are novel, and show how communication happens between iteration spaces.

Of these rules, (1) \textbf{LowerSum} is one of the two top-level entry rules. It replaces the high-level summation with an explicit accumulator variable $\tilde{C}_{(\vec{i}, r)}$ and initializes it to zero, producing a recurrence that iterates stepwise over $r$. For the matrix multiply example, this rule identifies that there is a summation over $k$ necessary to calculate $C_{ij} = \sum_{k} A_{ik}\cdot B_{kj}$, and then emits the recurrence $\tilC_{ij} \leftarrow (i,j,k-1)_{\tilC} + A_{ik} \cdot B_{kj} $

(2) \textbf{ChooseStream} is the other top level rule, being an entry point to compilation. It chooses a stream axis $s_t \in \vec{i}$ such that $s_t \notin \mathrm{indices}(X^{(t)})$. This ensures that a single injected value can be propagated along $s_t$ and reused across all dependent lattice points, thus establishing the direction of its propagation within the iteration space.  (3) \textbf{Inject} introduces the base of each stream: at the start of the chosen stream axis, we inject the operand’s source value $X^{(t)}_{g_t(\vec{i}, r)}$ into the iteration space. For the matrix multiply, this rule would identify $j$ as the stream for argument $A_{ik}$, as $j$ is not included in $A_{ik}'s$ indexing expression. It would similarly choose $i$ as the stream for argument $B_{kj}$. (4) \textbf{Propagate} defines how that value is received at subsequent lattice points, typically by shifting one coordinate:

\[
(\vec{i}, r)\, X^{(t)} \;\leftarrow\; (\vec{i}[s_t \mapsto s_t - 1], r)\, X^{(t)}.
\]
For the matrix multiply, this rule converts the input term $A_{ik}$ into $(i,j,k)_{A} \leftarrow (i,j\!-\!1,k)_{A}$, and then applies a similar transformation to $B_{kj}$. (5) \textbf{Send} expresses the dual forward movement, being the other side of propagate.

The outer send rule handle cases where there are non-distributive outer expressions around a summation. The inner send rule handles recurrences with dependencies, and dictates how an output gets reused as an input for further computations. For a triangular solve, the presence of an outer expression and output dependencies results in the boundary-handshake equations seen in Figure~\ref{fig:deps-tri-solve}. The triangular solve contains two operations, and thus two iteration spaces; one for the GEMM, and one for the smaller triangular solve, which can be visualized with nodes representing the dataflow in Figure~\ref{fig:trisSolveDataflow}

These rewrite rules show how a \emph{stream}() scheduling command gets lowered. The \emph{broadcast()}  command follows a slightly modified version of these rewrite rules; all streaming rules stay the same, except the propagate rule becomes irrelevant, and all inject/send instructions are from a source/feeder cell at the boundary of the PE grid, instead of from a neighboring processor.

Conceptually, these rewrites make the spatial temporal structure of each recurrence explicit. The resulting program operates not as a pure reduction but as a wavefront of local updates, each consuming values streamed from its neighbors and producing results for subsequent iterations.



\section{Compilation Targets}
Cyclotron compiles its input recurrences into a set of per-PE programs, each specifying the local computation and communication behavior of an individual processing element. These programs can be executed on one of two backend targets: a cycle-accurate simulator that models a lattice of chiplets on a single die, or a multinode backend that employs MPI for inter-PE communication across distributed systems. Adding a new target is straightforward: each backend implements a lightweight runtime in which every PE runs an interpreter that iterates through its instruction list and executes operations in order. The runtime need only provide primitives for sending and receiving messages, along with implementations of the arithmetic kernels used by the program. Both existing backends follow this model and remain compact—each under roughly a thousand lines of code—making it easy to extend Cyclotron to new architectures or communication substrates. This modular design decouples code generation from execution, allowing the same compiled program to run unmodified on both simulated and distributed environments.
\subsection{Simulator Target}

Cyclotron builds on the Dataflow Abstract Machine (DAM) ~\cite{zhang2024dataflow}, a cycle-accurate runtime for spatial architectures that models each PE as an independent \emph{context} executing its own instruction stream while communicating over bounded channels. DAM follows Communicating Sequential Processes with Time (CSPT): each context advances a local clock, exchanges timestamped messages, and synchronizes through lightweight atomic or futex-based primitives rather than a global event queue. This distributed notion of time enables accurate simulation of large systolic fabrics without centralized scheduling overhead.

A DAM configuration defines the grid topology of PEs (1D, 2D, or toroidal), the latency and bandwidth of each inter-PE link, and the capacity of each PE’s local register file, which limits how many intermediate values can be retained before backpressure occurs. Channels act as bounded FIFOs—if a producer sends into a full channel or a consumer reads from an empty one, execution naturally stalls, emulating the pipeline bubble and flow-control effects of real hardware. This makes DAM particularly well suited for modeling spatial accelerators where computation and communication interleave tightly.

Cyclotron lowers its per-PE systolic IR directly into DAM programs consisting of a minimal instruction set: \texttt{LOOP}, \texttt{SEND(chan, src)}, \texttt{RECV(chan, dst)}, and arithmetic kernels such as \texttt{GEMM}, \texttt{TRSM}, and \texttt{SOFTMAX}. Each PE executes its own sequence of these instructions while communicating with neighbors through explicit channels, preserving the original space–time dependencies encoded by the compiler’s schedule. The simulator executes all PE programs concurrently, collecting per-cycle metrics such as utilization, stall rates, and active instruction counts.

This backend grounds Cyclotron’s design-space exploration in hardware-realistic behavior. By sweeping DAM parameters such as link latency, FIFO bandwidth, register file size, and mesh topology, we can quantify how spatial mappings, stationary choices, and kernel fusions impact throughput and utilization.

\subsection{Multinode Target}
In addition to the DAM simulator, Cyclotron provides an MPI backend and runtime that executes the same per-PE programs across distributed memory systems. Each PE’s instruction stream is interpreted by the runtime, mirroring the structure of the DAM backend's instruction set and runtime. Each Cyclotron PE is mapped to an MPI rank, and each channel in the dataflow graph becomes an explicitly allocated communication buffer between ranks. The backend supports point-to-point primitives, including both blocking MPI calls and nonblocking MPI\_Isend and MPI\_Irecv pairs for data exchanges and the SUMMA-style broadcasts. Local compute kernels are implemented as BLAS or LAPACK calls. 

Unlike the DAM runtime which models cycle timing,the MPI backend executes at message granularity, prioritizing scalability with HPC clusters over cycle accuracy. This allows Cyclotron to run large-scale workloads on real systems. Our backend targets CPU clusters, and could be easily extendible to support GPU clusters as well, in which communication would simply use NCCL's send/receive primitives instead of MPI's. The DAM simulator and MPI backend ingest the exact same per-PE program emitted by the Cyclotron compiler. Together with the DAM simulator, the MPI backend provides a continuum from detailed hardware modeling to scalable execution, enabling end-to-end evaluation of systolic schedules—from simulated cycles to distributed performance on real clusters.
\section{Evaluation}

\begin{figure}
    \centering
    \includegraphics[width=0.5\linewidth]{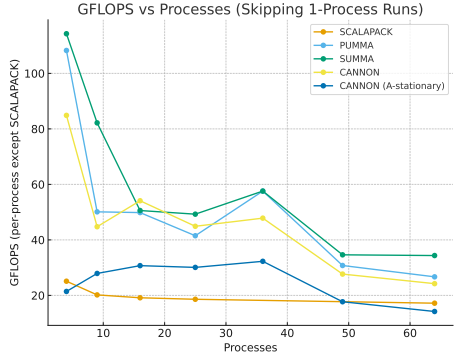}
    \caption{Cyclotron matrix multiply algorithms vs ScaLAPACK}
    \label{fig:matmulRes}
\end{figure}

\begin{figure}[t]
    \centering
    \begin{subfigure}[c]{0.4\linewidth}
        \centering
        \includegraphics[width=\linewidth]{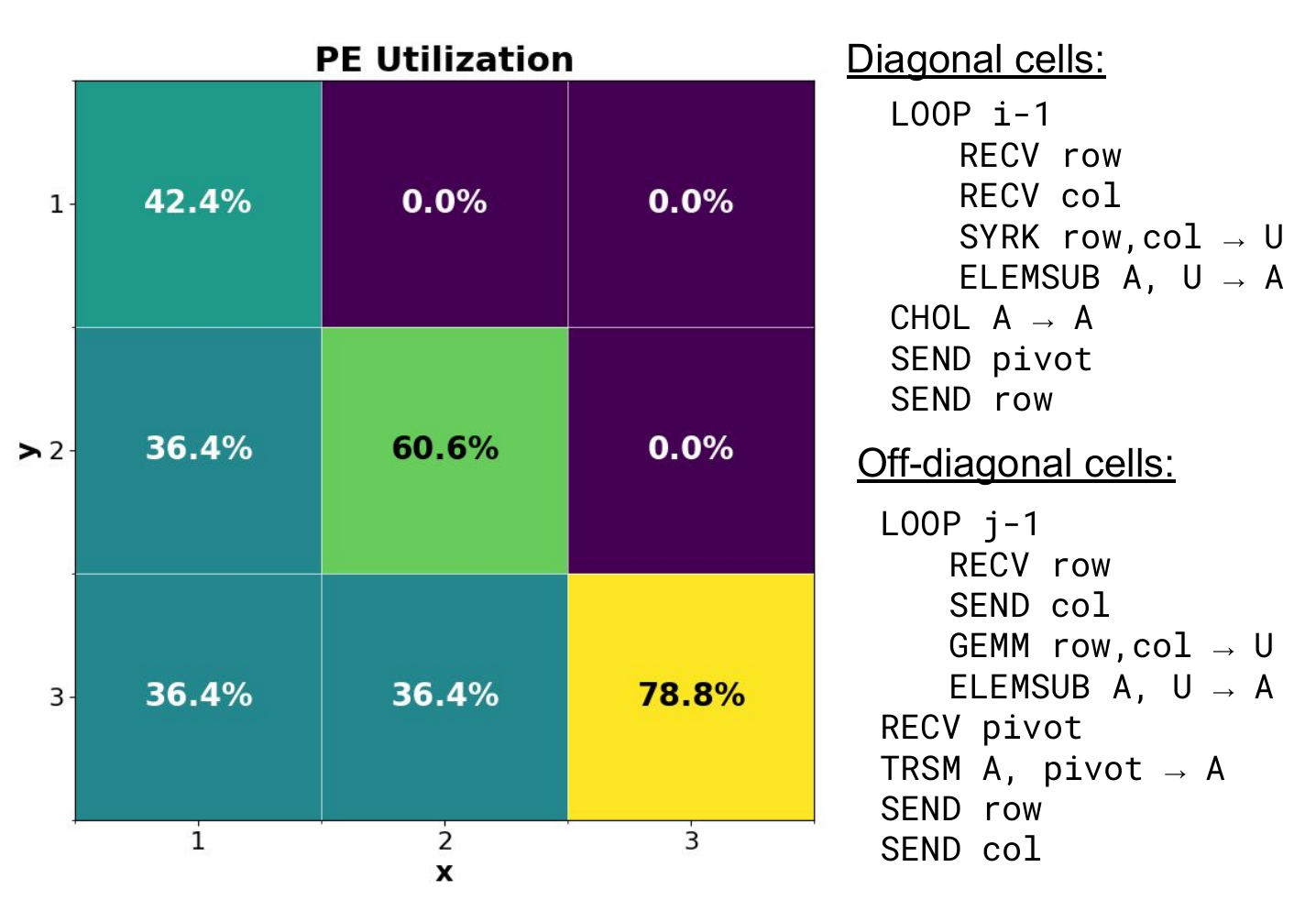}
        \caption{Cholesky Decomposition utilization under DAM backend.}
        \label{fig:trsmUT}
    \end{subfigure}
    \hfill
    \begin{subfigure}[c]{0.48\linewidth}
        \centering
        \includegraphics[width=\linewidth]{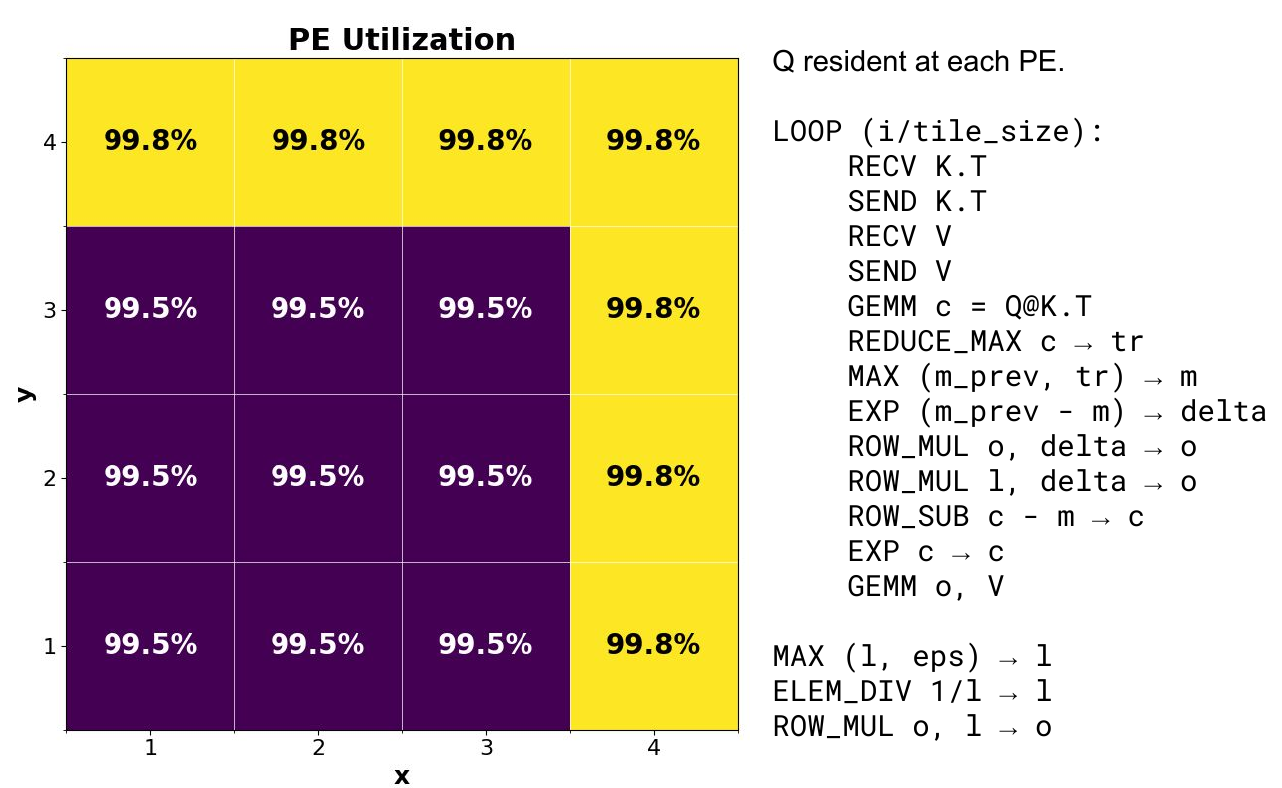}
        \caption{FlashAttention-2 PE activity and operation sequence.}
        \label{fig:fa2UT}
    \end{subfigure}
    \caption{(a) Example Cholesky mapping showing PE utilization. (b) FlashAttention-2 mapping with corresponding operation sequence.}
    \label{fig:pe-utilization-chol-fa2}
\end{figure}


\begin{figure}[t]
    \centering
    \begin{subfigure}[c]{0.58\linewidth}
        \centering
        \includegraphics[width=\linewidth]{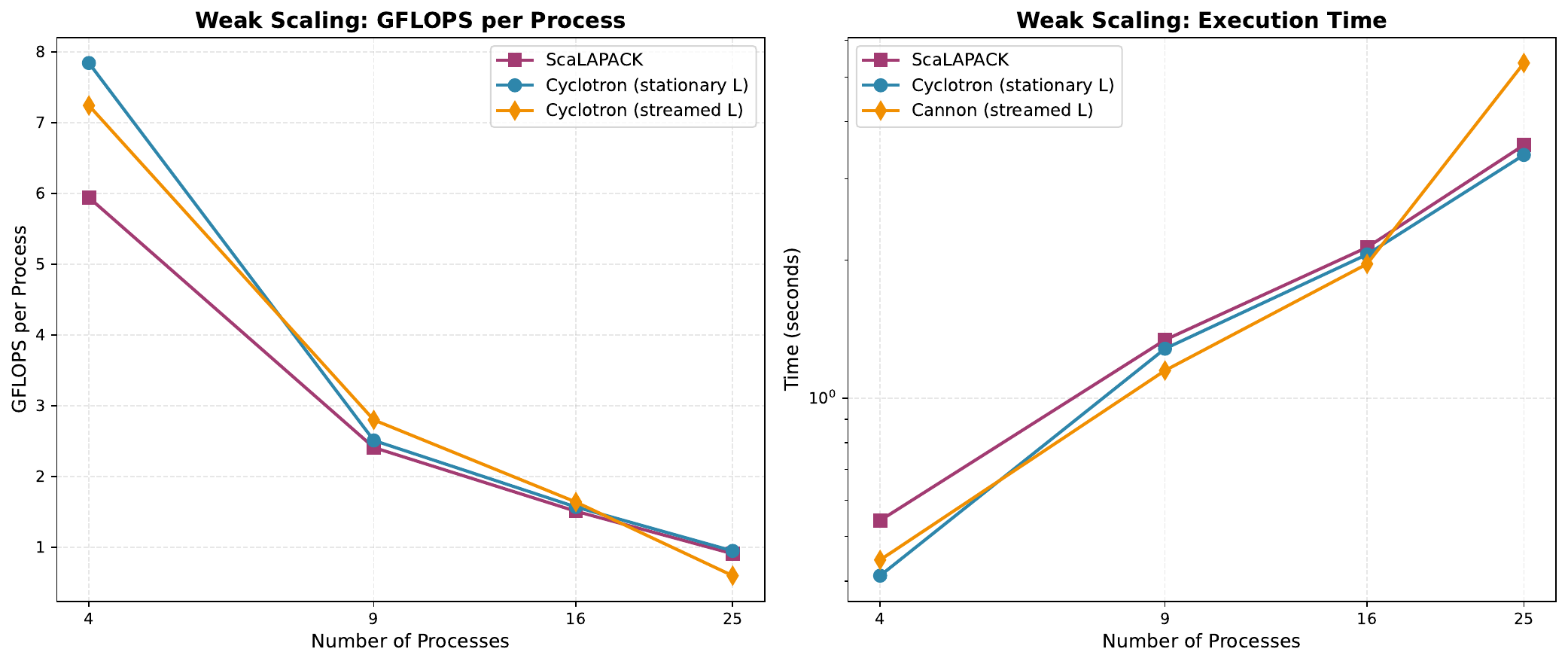}
        \caption{Weak scaling of TRSM compared to ScaLAPACK.}
        \label{fig:trsm-weak-scaling}
    \end{subfigure}
    \hfill
    \begin{subfigure}[c]{0.38\linewidth}
        \centering
        \includegraphics[width=\linewidth]{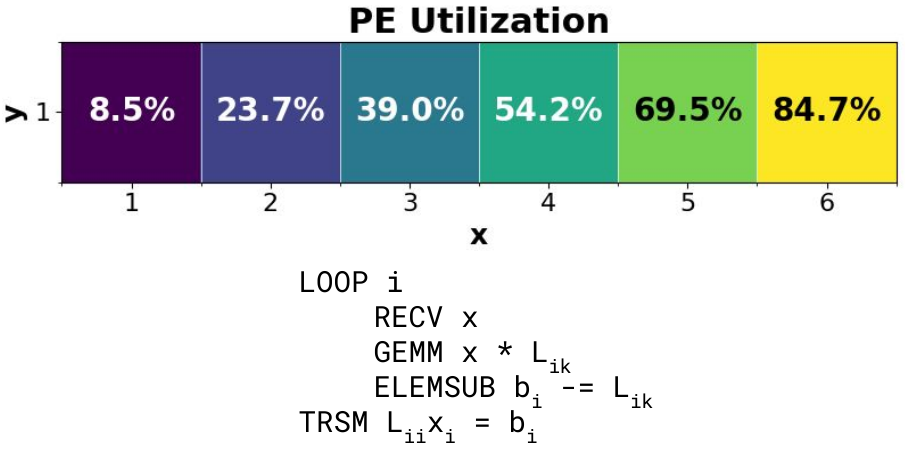}
        \caption{TRSM PE utilization under DAM backend.}
        \label{fig:trsm-dam-util}
    \end{subfigure}
    \caption{TRSM performance and utilization.}
    \label{fig:trsmUT}
\end{figure}

\subsection{Distributed Matrix Multiply}
We compare different implementations of distributed matrix multiply using our MPI-backend, comparing performance against ScaLAPACK, a standard handwritten library for distributed linear algebra (itself built on top of MPI). Using Cyclotron, we generate programs for output stationary Cannon's algorithm, weight stationary Cannon's algorithm, as well as PUMMA and SUMMA. Figure ~\ref{fig:matmulRes} shows that Cyclotron's performance generally outperforms ScaLAPACK, measured in GFLOPS/processor. We note that SUMMA, in which all processors receive all their inputs during an initial broadcast phase, performs best. It is the most asynchronous of the matrix multiply algorithms; processors do not wait on neighboring processors that are busy computing GEMMs, as they do in Cannon's algorithm. Because all data is initially broadcasted from the source nodes, the dependencies on neighboring processors that are inherent to Cannon's algorithm are now nonexistent. As PUMMA is an interpolation between Cannon's algorithm and SUMMA, it's performance is between the two. We note that in this case the weight stationary algorithm performs the slowest. We postulate this is due to the additional dependency weight stationary Cannon's algorithm has over output stationary Cannon's algorithm. In the weight stationary Cannon's algorithm, partial outputs $C$ must be computed by a GEMM, after which these $C$ blocks are sent to neighboring processors, representing a WRITE-READ dependency. In the output stationary algorithm, $C$ is kept local, and $A$ and $B$ shift across processors, meaning the dependency between the two can be relaxed, allowing the runtime to use asynchronous versions of the send and receive instructions.  We postulate performance improvement over ScaLAPACK is due to the much lighter runtime of Cyclotron vs. ScaLAPACK, as both use MPI in their communication layer and MKL-BLAS to perform local computations. 

\subsection{Distributed and Simulated Triangular Solve }

We show utilization results in DAM for triangular solve in Figure \ref{fig:trsm-dam-util}. As we can see, the utilization increases further down the array, since the number of \texttt{GEMM} updates grows longer as we progress further along $i$. We show multinode results for the triangular solve in Figure ~\ref{fig:trsm-weak-scaling}, in which we compare two schedules of the triangular solve against ScaLAPACK. In one schedule, $L$ is kept stationary with the \emph{prefetch} command, while $L$ is streamed in the other with the \emph{stream} command. All three implementations are comparable, showing the benefits of keeping an input matrix stationary in a triangular solve, whereas it did not help for the distributed matrix multiply.

\subsection{Simulated Cholesky and Flash Attention}

First, we introduce the tiled Cholesky decomposition and Flash Attention recurrences. 

\paragraph{Cholesky Recurrence.}
Let $A\in\mathbb{R}^{n\times n}$ be symmetric positive definite and partitioned into $b\times b$ tiles $A_{ij}$.
For each panel $j=1\ldots T$, the diagonal block accumulates all prior
Schur-complement updates via blocked \textsc{SYRK}:
\[
A_{jj}^{(j)} = A_{jj} - \sum_{k<j} \textsc{SYRK}(A_{jk}),
\qquad
L_{jj} = \textsc{CHOL}(A_{jj}^{(j)}).
\]
Each off-diagonal tile $i>j$ performs its corresponding update using blocked
\textsc{GEMM}, followed by a triangular solve:
\[
A_{ij}^{(j)} = A_{ij} - \sum_{k<j} \textsc{GEMM}(A_{ik}, A_{jk}),
\qquad
L_{ij} = \textsc{TRSM}(A_{ij}^{(j)}, L_{jj}).
\]
The resulting tiles $\{L_{ij}\}$ form the lower-triangular Cholesky factor.

\paragraph{FlashAttention-2 Recurrence.}
Let $Q\in\mathbb{R}^{N\times d}$, $K,V\in\mathbb{R}^{M\times d}$ and
$\alpha=1/\sqrt{d}$. Keys/values are processed in tiles $J_t$,
maintaining per-query state $(m_i^{(t)},\,l_i^{(t)},\,o_i^{(t)}\in\mathbb{R}^d)$:
\[
S_{i,J_t} = \alpha\, Q_{i,:} K_{J_t,:}^\top,\qquad
\hat{m}_i^{(t)} = \max\!\big(m_i^{(t-1)},\;\max S_{i,J_t}\big),
\]
\[
l_i^{(t)} = e^{m_i^{(t-1)}-\hat{m}_i^{(t)}}\, l_i^{(t-1)}
            + \sum_{j\in J_t} e^{\,S_{i,j}-\hat{m}_i^{(t)}},
\]
\[
o_i^{(t)} = e^{m_i^{(t-1)}-\hat{m}_i^{(t)}}\, o_i^{(t-1)}
            + \sum_{j\in J_t} e^{\,S_{i,j}-\hat{m}_i^{(t)}}\, V_{j,:},
\qquad
m_i^{(t)} = \hat{m}_i^{(t)}.
\]
Final outputs are $O_{i,:} = o_i^{(T)} / l_i^{(T)}$.

We compare the utilizations for these two recurrences in DAM, as shown in Figure \ref{fig:pe-utilization-chol-fa2}. Choleskey's tiled recurrence is triangular and panel-structured, yielding a wavefront of parallelism that expands down the column. Each diagonal cell performs a Cholesky decomposition on a tile, preceded by \texttt{SYRK} updates if necessary. These \texttt{SYRK} updates grow in length as the systolic array dimensions grow, leading to higher utilization on diagonal cells and higher utilization lower on the systolic array. In contrast, the Flash-Attention-2 recurrence is relatively balanced, as each PE holds a query tile and computes a series of matrix multiplications, followed by elementwise activations in place.



\section{Related Work}

Systolic Arrays were first introduced by Kung and Leiserson \citep{kung1979systolic,kung1982why}. Classic work by Rajopadhye and Fujimoto \citep{rajopadhye1990synthesizing} established how to synthesize systolic arrays from (uniform/affine) recurrence equations, including conditions for linear schedules and array allocation. Their framework is entirely theoretical, and does not include compiler infrastructure. Our treatment generalizes these foundations to modern multi-tensor kernels, making skews and inter-space interfaces first-class in the IR.

Our IR is an extension and re-imagining of the \emph{assignment free} notation from the\emph{Algorithms of Informatics} textbook by Gyires \citep{gyires_systolic}. Their notation is similar to Halide's input language \citep{ragan2013halide}, in which all logical access are array-style (e.g. $A(i,j,k)$, as opposed to our $(ijk)^{mul}_{A}$). This array notation is is shared by the Stellar \citep{genc2024stellar} and Gemmini \citep{genc2021gemmini} work for designing matrix-multiply systolic arrays. Our notation treats the indices and iteration space as the first-class citizens, and extends the notation to handle recurrences, multiple iteration spaces, and inter-space communication. The Wavefront Array Processor \citep{wavefrontLanguage} was a programmable systolic array. It included a language  that looked similar to the per-PE programs that Cyclotron generates. Early work on codesigning a processor and a systolic compiler (using the afine model) was done by \citet{lam2004data}. There is a long history of dataflow compilers based on the polyhedral and affine models. AutoSA \citep{wang2021autosa} and PolySA \citep{cong2018polysa} use the polyhedral model to find unit-length read distances amongst input equations, and then compile the equations to dataflow. However, they are primarily targeted at matrix multiply and convolutions, and do not handle recurrences (i.e., equations with dependencies) beyond the LU decomposition, as the LU decomposition having a similar dataflow to matrix multiply.

DISTAL \citep{distal}, the closest related project to this work, decouples tensor formats from distributed schedules, spanning both classic and modern matrix-multiplication algorithms (e.g., Cannon, COSMA) and compiling to a task runtime across CPU/GPU clusters. Its core contribution is a scheduling language that maps computation onto distributed machines, where operations “rotate’’ and move across processors. In contrast, our work focuses on the temporal and spatial alignment of operands—representing inter-processor data movement as streams within a unified recurrence abstraction rather than as discrete cluster-level tasks. Whereas DISTAL is primarily a tensor-algebra compiler and cannot express recurrences with output dependencies, our IR directly captures such dependencies and treats them as first-class citizens. Moreover, DISTAL supports only output-stationary algorithms, while our space–time mapping generalizes to input-stationary and mixed-stationary executions, enabling a broader class of distributed and systolic schedules.

The classic distributed matrix multiply algorithms implemented by Cyclotron and DISTAL include Cannon's algorithm \citep{cannon1969cellular}, SUMMA \citep{van1997summa}, and PUMMA \citep{choi1994pumma}. Early work by \citet{heathTriangular} explored different forms of the distributed triangular solve.

\section{Conclusion}
Cyclotron unifies the specification, scheduling, and execution of systolic algorithms within a single compiler framework. By expressing computations as recurrences over indexed tensors, it exposes both spatial and temporal structure directly in the IR, enabling the compiler to generate hardware-like schedules while remaining architecture-agnostic. Its per-PE compilation model and dual backends—one for simulation, the other distributed via MPI—bridge the gap between theoretical mappings and executable implementations. Through these abstractions, Cyclotron shows that high-level recurrence semantics can provide a practical foundation for reasoning about dataflow, synchronization, and communication across a wide range of architectures.

Looking ahead, Cyclotron opens several avenues for exploration: integrating automatic schedule synthesis, synthesizing hardware on reconfigurable architectures, and coupling the compiler with hardware cost models for co-design. More broadly, it points toward a unified programming model for spatial computing—spanning single-chip kernels to distributed accelerators—grounded in a single, recurrence-based view of computation.




\bibliographystyle{ACM-Reference-Format}
\bibliography{sample-base}

\appendix

\end{document}